\RequirePackage{fix-cm}
\documentclass{svjour3}                     
\usepackage[T1]{fontenc}
\usepackage[latin9]{inputenc}
\usepackage{amsmath}
\usepackage{amssymb}
\usepackage{graphicx}
\usepackage{subfig}
\usepackage[title]{appendix}
\usepackage{color}
\usepackage{microtype}

\usepackage{lineno} 
\newcommand{\DIF}{\textnormal{d}}
\journalname{Acta Geotechnica}
\begin{document}
%
\title{\Large\bfseries Stress-induced anisotropy in granular materials:\\
       fabric, stiffness, and permeability}
\titlerunning{Stress-induced anisotropy in granular materials}
\author{Matthew~R.~Kuhn \and WaiChing~Sun \and Qi~Wang}
\authorrunning{M. R. Kuhn et al.}
\institute{%
             M. R. Kuhn \at
             Department of Civil Engineering,
             Donald P. Shiley School of Engineering,
             University of Portland,
             5000 N. Willamette Blvd,
             Portland, OR 97203
             Tel: 503-943-7361
             Fax: 503-943-7316
             \email{kuhn@up.edu}
             \and
             W. Sun \and Q. Wang \at
             Department of Civil Engineering and Engineering Mechanics,
             The Fu Foundation School of Engineering and Applied Science,
             Columbia University in the City of New York,
             New York, NY  10027
          }
\date{Received: \today / Accepted: date}
\maketitle
\begin{abstract}
The loading of a granular material induces anisotropies of the particle
arrangement (fabric) and of the material's strength,
incremental stiffness, and
permeability.
Thirteen measures of fabric anisotropy are developed, which are arranged in
four categories:
as preferred orientations of the particle bodies, the particle
surfaces, the contact normals, and the void space.
Anisotropy of the voids is described through image analysis
and with Minkowski tensors.
The thirteen measures of anisotropy change during loading, as determined
with three-dimensional (3D) discrete element (DEM) simulations of biaxial
plane-strain compression with constant mean stress.
Assemblies with four different particle shapes were simulated.
The measures of contact orientation
are the most responsive to loading, and they change
greatly at small strains; whereas, the other measures lag the loading
process and continue to change beyond the state of peak
stress and even after the deviatoric stress has
nearly reached a steady state.
The paper implements a methodology for characterizing the incremental
stiffness of a granular assembly during biaxial loading,
with orthotropic loading increments
that preserve the principal axes
of the fabric and stiffness tensors.
The linear part of the
hypoplastic tangential stiffness is monitored
with oedometric loading increments.
This stiffness increases in the direction of the initial compressive
loading but decreases in the direction of extension.
Anisotropy of this stiffness is closely correlated with a particular
measure of the contact fabric.
Permeabilities are measured in three directions with
lattice Boltzmann methods at various stages of loading and for assemblies
with four particle shapes.
Effective permeability is negatively correlated with the directional mean free path
and is positively correlated with pore width,
indicating that the anisotropy of effective permeability induced
by loading is produced by changes in the directional
hydraulic radius.
\keywords{stress-induced anisotropy \and fabric \and granular material
\and anisotropic permeability \and discrete element method}
\end{abstract}
\section{Introduction}
Granular materials are known to exhibit a marked anisotropy of
mechanical and transport characteristics.
This anisotropy can be an inherent consequence
of the original manner in which the material
was assembled or deposited (i.e., the \emph{inherent anisotropy} that
was succinctly described by \cite{Arthur:1972a}), but the initial 
anisotropy is also altered by subsequent loading
(\emph{stress-induced anisotropy}, 
as in \cite{Arthur:1977b,Oda:1985a,Thornton:1990a}).
Anisotropy can be expressed as a mechanical stiffness or strength
that depends upon loading direction or as hydraulic, electrical, or
thermal conductivities that depend upon the direction of the
potential gradient.
Although the presence of anisotropy can
be directly detected as a preferential,
directional arrangement of grains,
it can also be subtly present in
the contact forces and contact stiffnesses
having a predominant orientation.
This direction-dependent character is often attributed to
the material's internal \emph{fabric}, a term that
usually connotes one of two meanings.
The fabric can be a measurable average of the micro-scale
particle arrangement
(such as the Satake contact tensor \cite{Satake:1982a}),
or it can be a conceptual
phenomenological quantity, often a tensor, that imparts
an anisotropic character to a continuum
constitutive description of the material (e.g. the fabric
tensor of Li and Dafalias \cite{Li:2002a}).
The current work attempts a bridge between these two views of
fabric and anisotropy, focusing on stress-induced anisotropy:
we identify those physically measurable micro-scale
attributes that are most closely correlated with the bulk
stress, stiffness, and permeability.
After a brief description of the DEM simulations that form the
basis of this study
(Section~\ref{sec:DEM}),
we catalog thirteen micro-scale measures of
fabric in Section~\ref{sec:fabric},
which are reckoned from the
biased orientations of the particle bodies and surfaces,
of the inter-particle contacts, and of the void space.
These measures are quantified for a suite of
simulated granular assemblies,
each with a different
particle shape, which are initially
isotropic but undergo stages of biaxial compression
that impart an induced anisotropy.
\par
Although fabric anisotropy is an established concept for
granular materials, the current work ascertains
correlations between specific fabric measures and
bulk anisotropies in the stress, stiffness, and permeability.
We also consider the associations among the thirteen
fabric measures and their relation to particle shape.
In Sections~\ref{sec:stiffness} and~\ref{sec:permeability},
we characterize the evolution of
the mechanical stiffness and the hydraulic conductivity.
By considering evolutions of the various fabric measures
and the consequent measured behaviors, we determine which of
the fabric measures of Section~\ref{sec:fabric}
are most closely associated with
anisotropies in stiffness and permeability.
\par
Stiffness anisotropy can be measured with stress probes
\cite{Calvetti:1997a,Hoque:1998a}
or by measuring p-wave speeds in different
directions \cite{Ishibashi:1991a,Santamarina:1996a}.
This anisotropy is known to be induced by several aspects of
granular loading.
During loading, contacts are depleted in directions of extension,
leading to a preponderance of contacts that are oriented in
the direction of compression loading 
\cite{Oda:1985a,Chen:1991a,Calvetti:1997a},
and because granular stiffness is largely derived from the stiffnesses
of contacts, any directional preponderance of the contacts promotes
an anisotropy of stiffness.
The most intensely loaded
contacts are usually spatially arranged in columnar
force chains that are more efficient in bearing stress
along their direction of orientation
\cite{Rothenburg:1989a,Majmudar:2005a}.
Among non-spherical elongated particles, the loading history
also tends to rotate
the particles so that their directions of elongation are perpendicular
to the direction of compression --- a direction that is favorable
to bearing further compression in this direction \cite{Oda:1972c,Oda:1985a}.
A more subtle anisotropy is induced in particles that interact through
Hertzian contacts.
Loading produces larger contact forces among those contacts that are
oriented in the direction of compression, and because the stiffness of a
Hertzian contact increases with increasing force,
a greater bulk stiffness is induced
in the direction of the compression
loading \cite{Wang:2008c}.
\par
The induced anisotropy is
not limited to the mechanical stress-strain relation.
The hydraulic properties of the granular assemblies, which
depend on both the size and geometrical features of the void space,
may also change due to re-arrangement of the voids and deformation
of the grain network.
From a theoretical standpoint, an anisotropy
of the permeability tensor means that it has eigenvalues
of distinct magnitudes, and there exist three orthogonal principal directions
corresponding to these eigenvalues
\cite{Chapuis1989,Bear2013}.
Wong \cite{Wong2003}
proposed a simple model that expands the Kozeny-Carman equation to an
anisotropic model by simply assuming that the strain and permeability tensors 
share the same principal directions.
This model is supported by experiments
on loose and dense sand specimens at low confining pressure.
The conclusion
is different than the one drawn in \cite{Zhu2007}, which considered
triaxial extension tests performed on porous sandstone
at confining pressures high enough to produce grain crushing.
In this case, the major principal direction of the permeability
aligns well with that of the major principal stress due to
induced microcracks that were preferentially aligned with
the maximum principal stress direction.
Sun et al. \cite{Sun2013} employed a lattice Boltzmann
model to directly compute effective permeabilities both along and orthogonal
to a shear band formed during simple shear loading.
This numerical
experiment suggests that anisotropic effective permeability
effects in granular assemblies
composed of spherical grains are not strong in the absence of grain
crushing.
A similar numerical approach will be used in this study.
\par
Flow through porous media depends on the porosity and on the size,
shape, and topology of the pore network, which have all received intense
interest in recent years
\cite{Fredrich1995,Wong2003,Sun2011a,Bear2013,Sun2015}.
The full characterization of sands and other
geomaterials can be
accomplished with thin-section and with non-invasive tomographic methods
analysis
\cite{Oda:1972a,DeHoff:1972a,Fredrich1995,Liang:2000a}.
In the former case, three-dimensional micro--structures are often statistically
reconstructed from one or multiple two-dimensional thin sections
\cite{Adler:1992a,Koutsourelakis2006,Zaretskiy2010b}. The effective
permeability of the micro--structures are then calculated by computational
fluid dynamics computer models or network models.
A drawback of this approach is that the inferred
effective permeability may depend on the quality of the pore geometry
reconstruction algorithm. Another approach is to directly calculate
the effective permeability from a 3D micro-CT image.
This method
has become increasingly popular in recent years due to the advancement
of micro-CT techniques, which both reduce the cost and improve the
resolution of micro-CT images.
Previous work, such as
\cite{Fredrich1995,Arns2005,White2006,Sun2011,Sun2011a},
has found that the estimated permeabilities inferred from
lattice Boltzmann and hybrid lattice Boltzmann-finite
element methods are consistent with experimental 
measurements, provided that the computational resolution is sufficient.
With both methods, the components of the anisotropic permeability tensor
can be calculated from a corresponding inverse problem for a given
specimen as shown in \cite{White2006}.
However, to study how evolution
of grain kinematics affects the anisotropic permeability,
volumetric digital image correlation (DIC)
techniques must be used on multiple
X-ray tomographic images such that the evolution of the grain fabric and 
pore geometry are both captured during the experimental test 
\cite{Hall:2010a, Ando:2013a}. 
While this experimental technique can provide invaluable micro-structural 
information at the grain scales, the execution of such a sophisticated 
experimental campaign that combines volumetric DIC
and X-ray tomographic imaging on a deforming specimen is not trivial.
As an alternative, Sun et al. \cite{Sun2013}
applied a region-growing method on
a deforming granular assemblies to
obtain micro--structures from discrete element simulations.
The advantage of this approach is that
one can study the interconnection between
the grain kinematics and hydraulic
properties for identical micro-structures subjected
to different loading paths,
without worrying about the difficulty of preparing identical 
physical specimens in a laboratory setting.
This approach has been adopted in this study.
%
\section{DEM Simulations}\label{sec:DEM}
\begin{figure}
  \centering\includegraphics[scale=0.16]{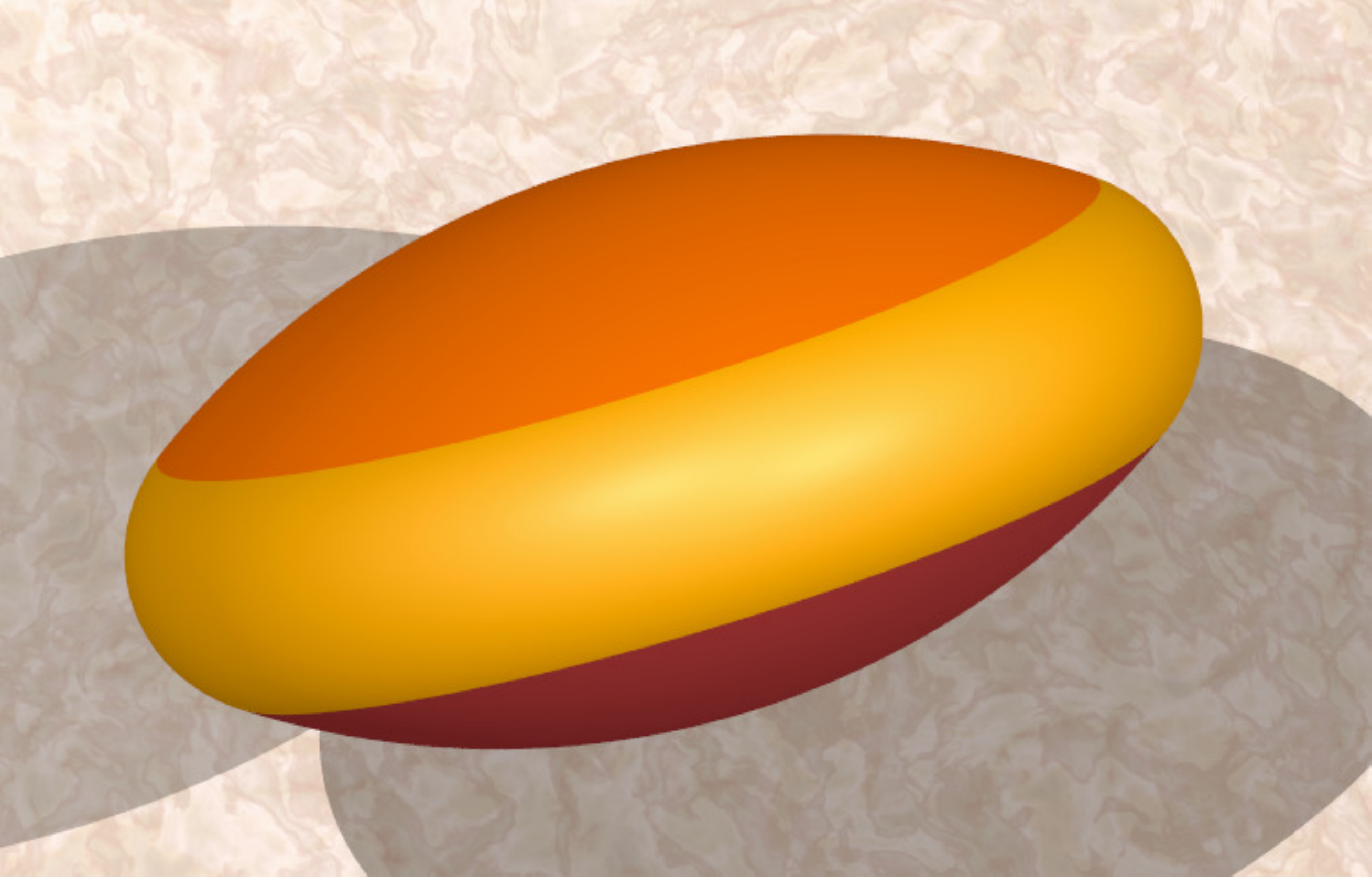}
  \caption{Ovoid particle, $\alpha=0.50$.
  \label{fig:Ovoid}}
\end{figure}

\begin{figure}
  \centering\includegraphics[scale=0.4]{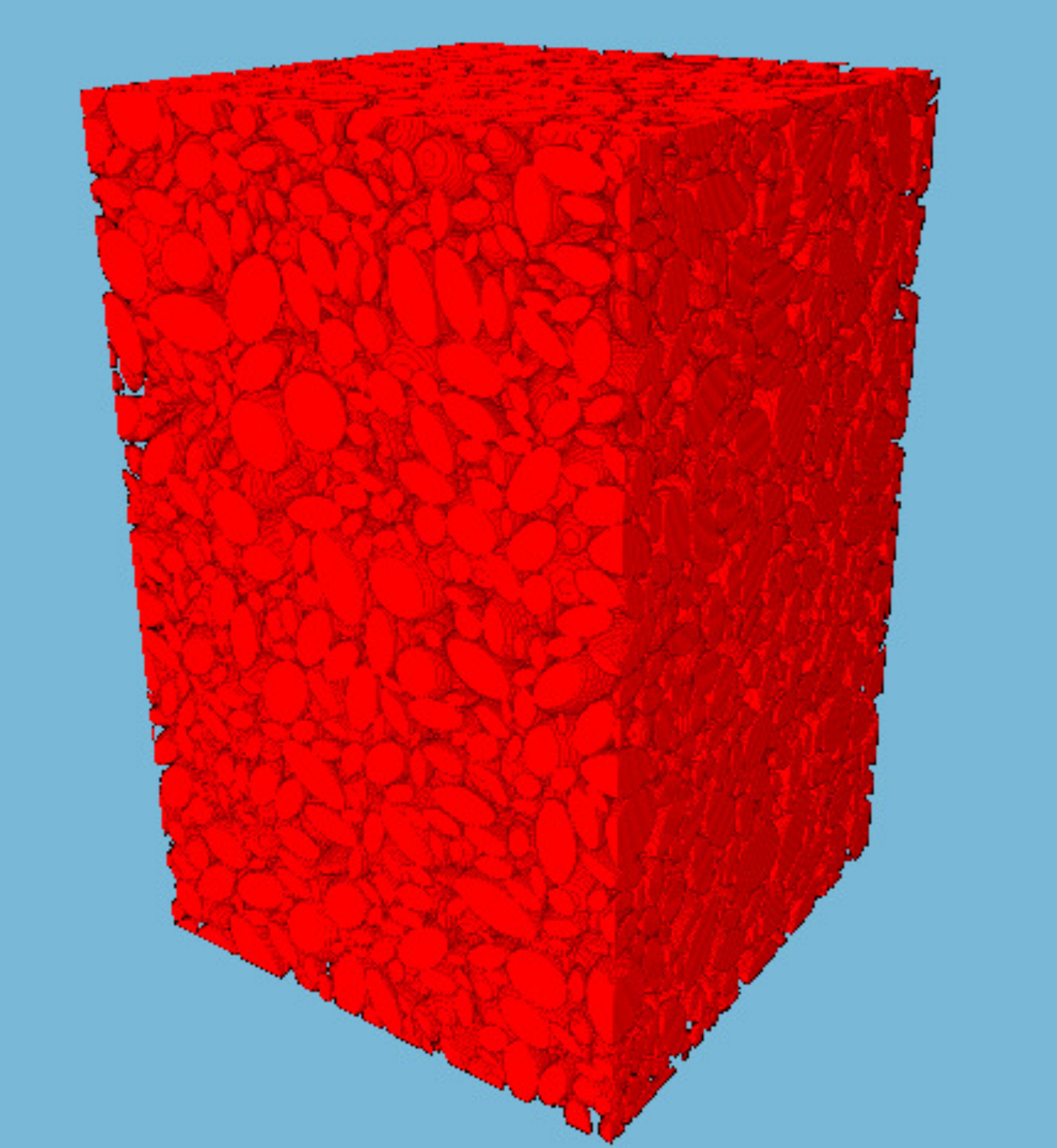}
  \caption{Dense initial assembly of 6400 ovoid particles, $\alpha=0.50$.
           \label{fig:Assembly}}
\end{figure}

Stress-induced anisotropy was investigated with four initially isotropic
DEM assemblies of unbonded smooth particles: one assembly of spheres
and three assemblies of oblate ovoid shapes having different aspect
ratios.
An ovoid is a convex composite solid of revolution with a
central torus and two spherical caps, approximating an oblate (flattened)
spheroid but allowing rapid contact detection 
and force resolution (Fig.~\ref{fig:Ovoid})
\cite{Kuhn:2003a}.
The four assemblies were composed of particles having aspect ratios
$\alpha$ of 1.0 (spheres), 0.80, 0.625, and 0.50, with sphericity decreasing
with lower $\alpha$.
The DEM simulations in this
study are element tests in which small assemblies of ``particles in
a box'' undergo biaxial plane strain compression. 
The purpose is to
explore the material behavior of a simulated granular element --- at both
macro- and micro-scales --- rather than to study a larger boundary
value problem that would require, perhaps, many millions of particles.
Figure~\ref{fig:Assembly} shows an initial, unloaded assembly of
6400 ovoid particles with aspect ratio 0.500, representing
a small soil element of size $18\times12\times12D_{50}$: an assembly
large enough to capture the average material behavior but sufficiently
small to suppress meso-scale localization in the form of shear bands.
The particles in this image are truncated along the flat periodic
boundaries that surround the assembly. 
The median particle size (by
volume) $D_{50}$ was 0.17mm for all four assemblies, with a size
range of 0.075mm to 0.28mm. 
For the ovoid particles, these sizes are
mid-plane diameters, with the axial height being smaller by factors
$\alpha$ of 0.80, 0.625, and 0.50 compared with the mid-plane diameter.
\par
The open-source OVAL code was used for the simulations \cite{Kuhn:2002b}.
The particles were isotropically compacted into fairly dense arrangements.
To construct these arrangements, the 6400 particles were sparsely
and randomly arranged within a spatial cell surrounded by periodic
boundaries. 
In the absence of gravity and with a reduced inter-particle
friction coefficient ($\mu=0.30$), the assembly was slowly and isotropically
compacted by reducing the dimensions in all directions by equal ratios.
The initially sparse arrangement with zero stress would eventually
``seize'' when a loose, yet load-bearing, fabric had formed. 
A series
of fourteen progressively denser assemblies was created by repeatedly
assigning random velocities to particles of the previous assembly
(simulating a disturbed or vibrated state) and then further reducing
the assembly volume until the newer specimen had seized. 
This compaction
procedure, when applied to mono-disperse spheres, results in isotropic
assemblies with a range of void ratios that compares favorably with
the range found with glass ballotini \cite{Kuhn:2014c}. 
For the current study, we selected
four assemblies having about the same
void ratio $e=0.556$ (porosity $n=0.358$):
one assembly of spheres and one assembly of each ovoid shape. 
After creating these dense assemblies, they were isotropically 
consolidated
to a common mean confinement stress of $p_{\text{{o}}}=100$kPa. 
\par
During subsequent deviatoric loading, the following properties were
assigned to the particles: shear modulus $G=29$GPa, Poisson ratio
$\nu=0.15$, and inter-particle friction coefficient $\mu=0.50$.
These values are in the ranges of those measured for quartz grains
\cite{Mitchell:2005a}. 
Hertz-Mindlin interactions were assumed in
computing the normal and tangential forces, and no contact bonding
or contact rotational resistance was included in the simulations. 
\par
The same slow, quasi-static loading conditions were applied to all
four assemblies: biaxial plane strain
compression with constant mean stress. 
To load the assemblies, the
larger dimension in the $x_{1}$ direction was reduced at a constant
rate (vertical direction in Fig.~\ref{fig:Assembly}), 
while maintaining constant width in the $x_{2}$ direction.
The width in the $x_{3}$ direction was continually adjusted so that
the mean stress $p$ remained 100kPa. 
Because of the periodic boundaries, no gravity was applied
in the simulations.
The containing periodic
cell remained rectangular, so that the directions of principal stress
remained aligned with the assembly boundaries.
Such loading conditions
would be expected to produce an orthotropic fabric, with the three
principal directions of fabric anisotropy aligned with the principal
stress directions.
\par
\begin{figure}
  \centering
  \mbox{%
    \subfloat[]{\includegraphics{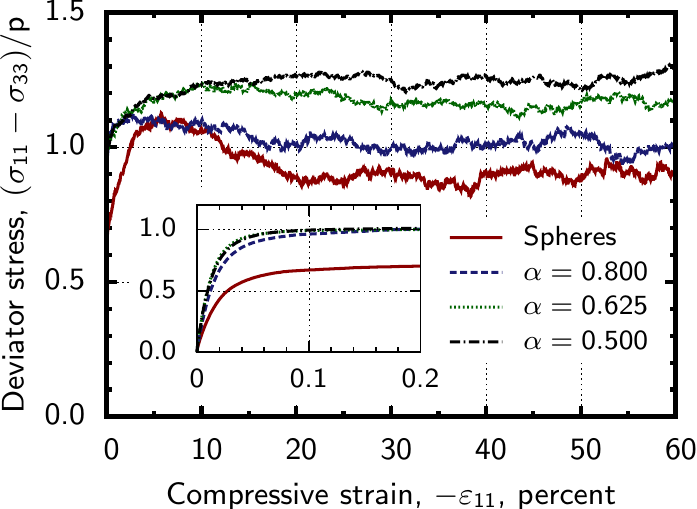}}
    \quad
    \subfloat[]{\includegraphics{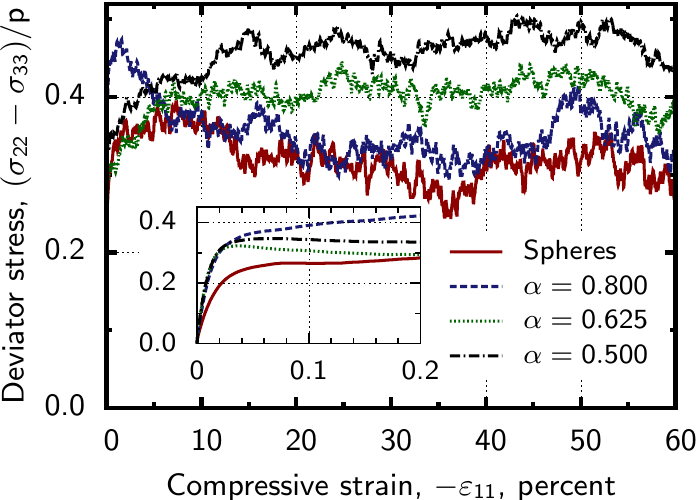}}
  }\\[1ex]
  \mbox{%
    \subfloat[]{\includegraphics{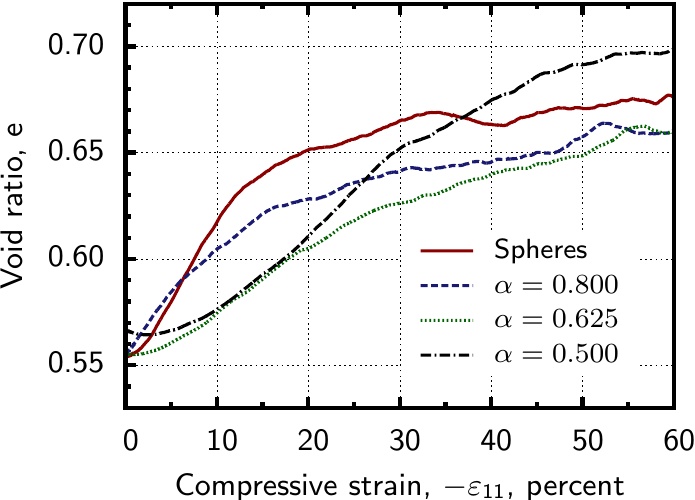}}
    \quad
    \subfloat[]{\includegraphics{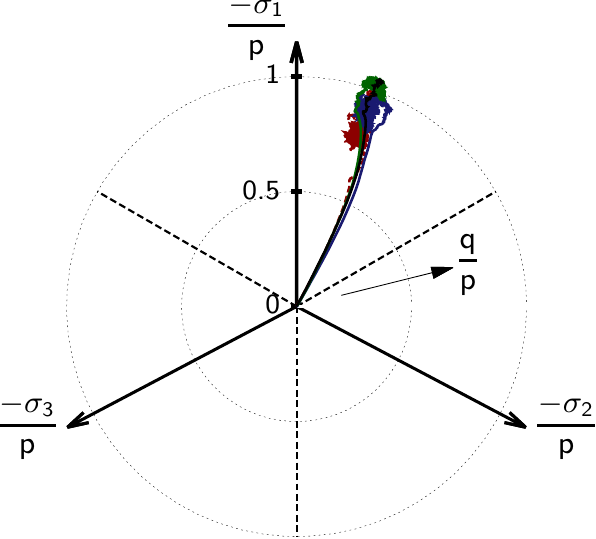}}
  }
  \caption{Mechanical response during biaxial plane strain compression:
           (a) deviatoric stress $\sigma_{11}-\sigma_{33}$,
           (b) intermediate deviatoric stress $\sigma_{22}-\sigma_{33}$
           (c) volume change, and 
           (d) $\pi$-plane stress paths, in which
           the radial scale is $q/p$, where $q=\sqrt{3J_{2}}$ 
           with second principal invariant 
           $J_{2}=[(\sigma_{1}-\sigma_{2})^{2}+(\sigma_{1}-\sigma_{3})^{2}+(\sigma_{2}-\sigma_{3})^{2}]/6$.
           Inset plots detail the small-strain behavior.
           \label{fig:Stress_strain}}
\end{figure}

Figure~\ref{fig:Stress_strain}a shows the deviatoric stress ratio
$(\sigma_{11}-\sigma_{33})/p$ that results from compressive strain
$-\varepsilon_{11}$.
Strength is seen to increase with decreasing sphericity, and the
assembly with the flattest particles ($\alpha=0.50$) had the
greatest strength.
This trend is consistent with experiments
\cite{Konishi:1982a,Oda:1985a}
and with other 
simulations~\cite{Rothenburg:1992b,Pena:2007a,Azema:2007a}.
Evolution of the intermediate principal stress is shown
in Fig.~\ref{fig:Stress_strain}b.
In both Figs.~\ref{fig:Stress_strain}a and~\ref{fig:Stress_strain}b,
the small-strain behavior is detailed in the smaller inset plots.
The volume change, shown in Fig.~\ref{fig:Stress_strain}c, follows
a less consistent trend. 
All assemblies began with about the same
void ratio, and all assemblies underwent extensive dilation after
the strain exceeded 5\%. 
That is, all assemblies were initially
dense relative to the critical state,
as all dilated significantly during the deviatoric loading.
Although a strain
of 60\% was insufficient to bring non-sphere specimens to a critical state
of isochoric plastic flow, the least-spherical ovoids ($\alpha=0.50$) attained
the largest void ratio, followed by the spheres and the ovoids with
$\alpha$'s of 0.80 and 0.625. 
The $\pi$-plane in Fig.~\ref{fig:Stress_strain}d
shows the evolution of the intermediate principal stress $\sigma_{22}$
during loading with constant $p$. 
Although the stress-strain and
volume change behavior are quite different for the four particle shapes
(Figs.~\ref{fig:Stress_strain}a and \ref{fig:Stress_strain}c), all shapes
share nearly the same path of principal stress evolution.
\par
Complete assembly information was collected
at numerous strains --- as small as 0.002\% and as large as 60\% --- during
the biaxial compression, so that loading-induced
anisotropies of fabric, stiffness,
and permeability could be measured.
These characteristics
are described in the
Sections~\ref{sec:fabric}, \ref{sec:stiffness}, 
and \ref{sec:permeability}, in which
we determine correlations among the different fabric measures
and between the fabric and the bulk stiffness and permeability.
\section{Evolution of fabric anisotropy}\label{sec:fabric}
In this section, we catalog several fabric measures
and describe their changes induced by biaxial compression loading.
The fabric measures
are listed in Table~\ref{tab:Measures}, which 
places them in four categories, depending on the object of focus:
particles, particle surfaces, contacts between particles, and
void space.
Many of the measures are either second-order tensors or
are rendered as matrices,
as are appropriate for characterizing anisotropy
(as a counter-example, void ratio and density are
scalar quantities and are inappropriate for portraying anisotropy).
As will be seen, the four categories are not clearly distinct, and some
measures are associated with multiple objects.
\begin{table}
  \caption{Fabric measures.\label{tab:Measures}}
  \centering\small%
  \begin{tabular}{lllll}
    \hline 
    Object & Section & Measure & Eqs. & Description\\
    \hline
    Particles, ``p'' & Sec.~\ref{sec:bodies} & $\overline{\mathbf{J}}^{\,\text{p}}$ & (\ref{eq:Jbari}), (\ref{eq:Jbar}) & Particle axes orientation tensor\\
    \hline 
    Surfaces, ``s'' & Sec.~\ref{sec:surfaces} & $\overline{\mathbf{I}}^{\,\text{s}}$ & (\ref{eq:Ibari}), (\ref{eq:Ipar}) & Surface inertia tensor\\
     && $\overline{\mathbf{S}}^{\,\text{s}}$ & (\ref{eq:Sbar}) & Surface normals orientation tensor\\
    \hline 
    Contacts, ``c'' & Sec.~\ref{sec:contacts} & $\overline{\mathbf{F}}^{\,\text{c}}$, $\overline{\mathbf{F}}{}^{\,\text{c-strong}}$ & (\ref{eq:Fc}), (\ref{eq:Fcstrong}) & Contact normal orientation tensors\\
     && $\overline{\mathbf{G}}^{\,\text{c}}$, $\overline{\mathbf{G}}{}^{\,\text{c-strong}}$ & (\ref{eq:Lc}) & Branch vector orientation tensors\\
     && $\overline{\mathbf{H}}^{\,\text{c}}$, $\overline{\mathbf{H}}{}^{\,\text{c-strong}}$ & (\ref{eq:Hbarc}) & Mixed-vector orientation tensors\\
    \hline
    Voids, ``v'' & Sec.~\ref{sec:voids} & $f_{\rho}^{\text{v}}(\rho )$ & (\ref{eq:fr}) & Distribution: void sizes\\
     && $f_{\ell_{i}}^{\text{v}}(\ell_{i})$ & (\ref{eq:fli}) & Distribution: directed free paths\\
     && $f_{r_{i}}^{\text{v}}(r_{i})$ & (\ref{eq:fri}) & Distribution: directed radial breadths\\
     && $\overline{\mathbf{L}}^{\,\text{v}}$ & (\ref{eq:Lambdaii}) & Matrix: median void free path\\
     && $\overline{\mathbf{R}}^{\,\text{v}}$ & (\ref{eq:Rii}) & Matrix: median void radial breadth\\
     && $\overline{\chi}^{\,\text{v}}$ & (\ref{eq:Chis1})--(\ref{eq:Chivbar}) & Scalar: void connectivity per particle \\
     && $\overline{\mathbf{W}}_{3}^{\,\text{v},2,0}$ & (\ref{eq:W202}), (\ref{eq:W320}), (\ref{eq:MinkNormal}) & Tensor: void shape and connectivity\\
    \hline 
  \end{tabular}
\end{table}
\subsection{Particle bodies}\label{sec:bodies}
The simplest (and most apparent) 
measures of fabric anisotropy are those based
upon the orientations of the particles --- information
that can be gathered from digitized images of physical specimens,
from the geometric data of computer simulations,
or by simply disassembling a physical packing of grains.
One such measure
addresses the orientations of elongated or flat particles.
For example,
Oda \cite{Oda:1972c}
presented histograms of particle orientation as measured from optical
micro--graphs of sheared sand specimens.
If the particles are nearly ellipsoidal, then we can identify
the three orthogonal directions of an ``$i\,$th'' particle's principal
axes (the unit column vectors $\mathbf{q}_{1}^{\text{p},i}$) and
the particle's corresponding widths 
in these directions (widths $a_{1}^{\text{p},i}$),
where the superscript ``p'' will denote ``particle'' information. 
The information for the single
particle can then be collected in a diagonal matrix
$\mathbf{A}^{\text{p},i}$ and in a matrix $\mathbf{J}^{\text{p},i}$
comprised of orthogonal column vectors $\mathbf{q}_{k}^{\text{p},i}$: 
\begin{gather}
  \mathbf{A}^{\text{p},i}=\left[\begin{array}{ccc}
  a_{1}^{\text{p},i} & 0 & 0\\
  0 & a_{2}^{\text{p},i} & 0\\
  0 & 0 & a_{3}^{\text{p},i}
  \end{array}\right]\\
  \mathbf{J}^{\text{p},i}=
  \left[  \mathbf{q}_{1}^{\text{p},i},
        \:\mathbf{q}_{2}^{\text{p},i},
        \:\mathbf{q}_{3}^{\text{p},i}\right]
  \label{eq:Jbari}
\end{gather}
with $\mathbf{A}^{\text{p},i}$ providing the widths
and $\mathbf{J}^{\text{p},i}$ providing the orientations.
For the ovoid shapes of our simulations, $\mathbf{q}_{1}^{\text{p},i}$
is oriented along a particle's central axis,
and the remaining orthogonal vectors,
$\mathbf{q}_{2}^{\text{p},i}$ and $\mathbf{q}_{3}^{\text{p},i}$,
are oriented in arbitrary transverse directions.
When averaged among all particles in an assembly or image, these quantities
can be used to compute a tensor-valued measure of the average particle
orientation: 
\begin{equation}
  \overline{\mathbf{J}}^{\,\text{p}}=
  \left\langle \frac{3}{\text{tr}\left(\mathbf{A}^{\text{p},i}\right)}\,
  J_{jm}^{\text{p},i}A_{ml}^{\text{p},i}J_{kl}^{\text{p},i}\right\rangle 
  \mathbf{e}_{j}\otimes\mathbf{e}_{k}
  \label{eq:Jbar}
\end{equation}
where brackets $\langle\circ\rangle$ designate an average
(in this case, of all particles $p$), and the $\mathbf{e}$
are Cartesian basis vectors.
Tensor $\overline{\mathbf{J}}^{\text{p}}$ is similar to the orientation
tensor of Oda \cite{Oda:1985a} but includes a factor that eliminates
the bias of particle size, so that each particle is given an equal
weight, regardless of its size. 
(The factors $3/\text{tr}(\mathbf{A}^{\text{p},i})$
can be removed if a size-bias is desired.) 
The factor also normalizes
the tensor, so that an assembly of spheres would yield the identity
(Kronecker)
tensor for $\overline{\mathbf{J}}{}^{\,\text{p}}$.
\par
\begin{figure}
  \centering\includegraphics{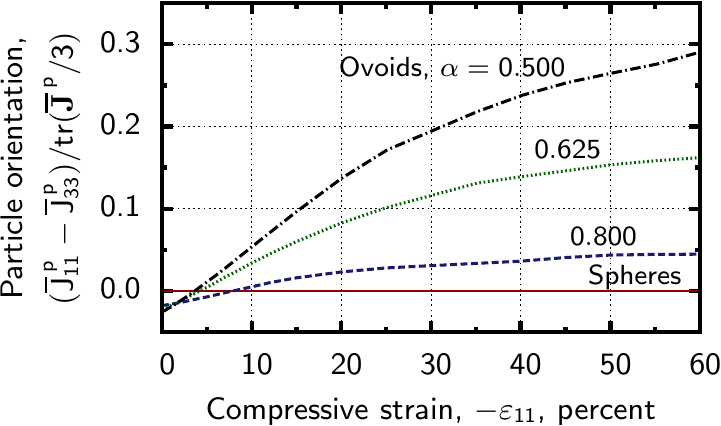}
  \caption{Anisotropy of the particle orientations.
           Compressive loading is in the $x_{1}$ direction.
           \label{fig:J_bar_p}}
\end{figure}

The evolution of anisotropy of particle orientations is shown in 
Fig.~\ref{fig:J_bar_p}
for the four assemblies during biaxial plane strain compression.
The figure gives the deviatoric difference 
$\overline{J}_{11}^{\,\text{p}}-\overline{J}_{33}^{\,\text{p}}$
divided by the mean value of particle orientation 
tensor $\overline{\mathbf{J}}^{\,\text{p}}$ 
(i.e., $1/3$ of the trace $\text{tr}(\overline{\mathbf{J}}^{\,\text{p}})$).
The difference is always 0 for sphere assemblies.
The ovoid assemblies
all begin with a slight anisotropy, but loading results in particles
whose wider dimensions are predominantly aligned with the extension
direction $x_{3}$, a phenomenon that is widely reported in both 2D
\cite{Rothenburg:1992b,Azema:2007a}
and 3D simulations \cite{Ouadfel:1999a,Ng:2001a}
and in physical experiments with sands and assemblies of plastic particles
\cite{Oda:1972c,Oda:1985a}. 
This tendency is apparent in
Fig.~\ref{fig:Digitized} for flattened ovoids ($\alpha=0.6262$)
at strain $-\varepsilon_{11}=60\%$. 
Figure~\ref{fig:J_bar_p} shows
that particle reorientation steadily progresses across the full range
of strains, even after the stress condition has nearly reached a steady-state
condition (compare with Fig.~\ref{fig:Stress_strain}a). 
With assemblies
of the flattest ovoids, a stationary fabric of particle orientation
is not yet attained at the largest strain of 60\%. 
At small strains,
very little reorientation occurs, even as the deviatoric stress
approaches its peak condition: 
the ratio $(\overline{J}_{11}^{\,\text{p}}-\overline{J}_{33}^{\,\text{p}})/(\text{tr}(\overline{\mathbf{J}}^{\,\text{p}})/3)$
has changed by less than 0.02\% at the strain $-\varepsilon_{11}=0.02$,
which is roughly at the peak stress 
state: further deformation is required to attain a steady state
of fabric.
These results show that particle reorientation
lags changes in stress during the early stage of loading, but
fabric change is more prolonged, and a
critical state (steady state)  of fabric
is not necessarily attained when only the volume and stress
are constant.
%
%
\subsection{Particle surfaces}\label{sec:surfaces}
A set of anisotropy
measures are associated with the orientations of particle
surfaces. The inertia tensor of a particle's surface is
%
\begin{equation}
  \mathbf{I}^{\text{s},i}=\int_{\partial S^{i}}\mathbf{x}
  \otimes\mathbf{x}\, \DIF A^{i}
  \label{eq:Ibari}
\end{equation}
where $\mathbf{x}$ is the vector from the centroid of the
particle's surface
$\partial S^{i}$ to points on this surface,
and $\mathbf{x}\otimes\mathbf{x}$ is the dyad $x_{i}x_{j}$. 
Superscript ``s'' denotes
a surface quantity of the particles. 
The average among all particles is 
\begin{equation}
  \overline{\mathbf{I}}^{\,\text{s}}=
  \left\langle \frac{3}{\text{tr}\left(\mathbf{I}^{\text{s},i}\right)}\,
  \mathbf{I}^{\text{s},i}\right\rangle 
  \label{eq:Ipar}
\end{equation}
which has been normalized in a similar manner as 
$\overline{\mathbf{J}}{}^{\,\text{p}}$ (Eq.~\ref{eq:Jbar}).
\par
Kuo et al. \cite{Kuo:1998a} introduced an orientation measure by
approximating the particle surface area (per unit of volume) of the
particles in direction $\mathbf{n}$,
\begin{equation}
  S(\mathbf{n})\approx
  \frac{S_{\text{v}}}{4\pi V}
  \left(1+\overline{Q}_{ij}^{\,\text{s}}n_{i}n_{j}\right)
\end{equation}
where $S(\mathbf{n})$ is a distribution function, 
$\overline{\mathbf{Q}}^{\,\text{s}}$
is a surface area tensor, and $S_{\text{v}}$ is the total surface
area of particles in volume $V$. 
They used stereological methods
proposed by Kanatani \cite{Kanatani:1988a} to estimate $\overline{\mathbf{S}}^{\,\text{s}}$
from 2D images along three orthogonal planes.
With DEM geometric data,
we can directly compute a similar average 
orientation tensor $\overline{\mathbf{S}}^{\,\text{s}}$
by integrating the dyads $n_{i}n_{j}$ across the surface of each of
the $N_{\text{p}}$ particles:
\begin{equation}
  \overline{\mathbf{S}}^{\,\text{s}}
  =\frac{3}{S_{\text{v}}}
  \sum_{i=1}^{N_{\text{p}}}
  \int_{\partial S^{i}}\mathbf{n}\otimes\mathbf{n}\, \DIF A^{i}
  \label{eq:Sbar}
\end{equation}
in which $S_{\text{v}}$ is the total area $\sum\int\, \DIF A$. 
In this
definition, the tensor is normalized 
so that $\overline{\mathbf{S}}^{\,\text{s}}$
is the identity matrix for an assembly of spheres.
If we divide by the total mass instead of by $S_{\text{v}}$,
Eq.~(\ref{eq:Sbar}) yields a corresponding measure of
specific surface that incorporates its anisotropic character.
\par
\begin{figure}
  \centering
  \mbox{%
    \subfloat[]{\includegraphics[width=0.485\textwidth]{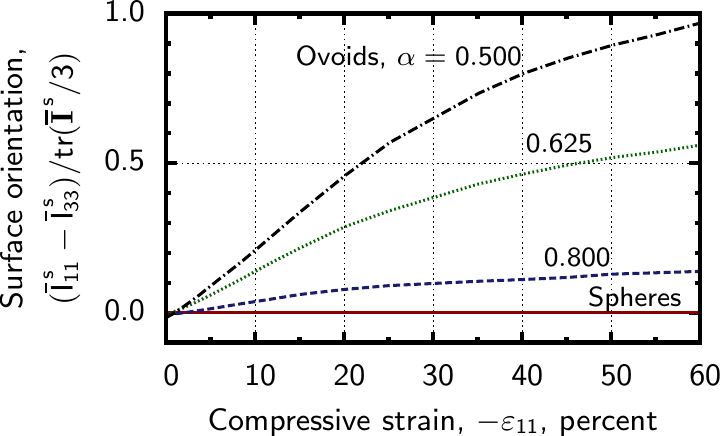}}
    \quad
    \subfloat[]{\includegraphics[width=0.485\textwidth]{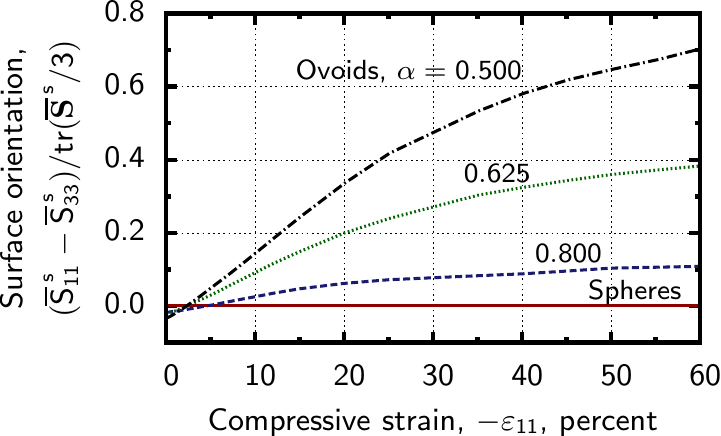}}
  }\\
  \includegraphics[width=0.485\textwidth]{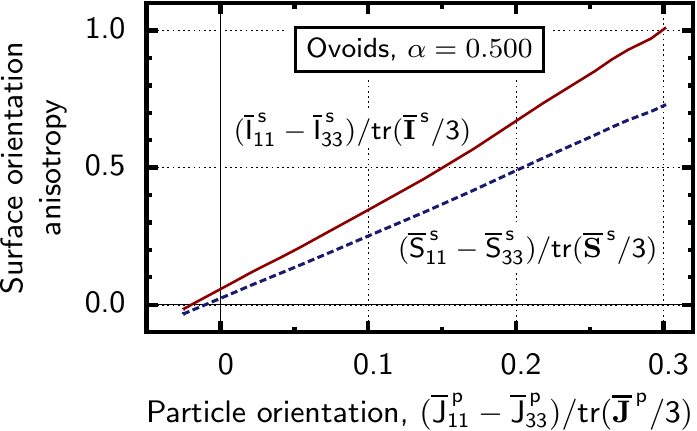}
  \caption{Anisotropy of particle surface orientations
           during biaxial compression:
           (a) orientation $\overline{\mathbf{I}}^{\text{s}}$,
           (b) orientation $\overline{\mathbf{S}}^{\text{s}}$,
           and (c) correlations between the anisotropies
           of surface orientation and of particle orientation for
           ovoids with $\alpha=0.500$.
  \label{fig:Surfaces_vs_strain}}
\end{figure}

The evolution of $\overline{\mathbf{I}}^{\,\text{s}}$ 
and $\overline{\mathbf{S}}^{\,\text{s}}$
during biaxial compression is shown in Figs.~\ref{fig:Surfaces_vs_strain}a
and~\ref{fig:Surfaces_vs_strain}b
for the four shapes.
The same trends in both measures are
similar to those in the previous
Fig.~\ref{fig:J_bar_p} of the particle orientation
tensor $\overline{\mathbf{J}}^{\,\text{p}}$ 
(correlations are shown in Fig.~\ref{fig:Surfaces_vs_strain}c):
(1)~the anisotropy of the particle surfaces increases with increasing
non-sphericity of the particle shape (smaller $\alpha$), 
(2)~anisotropy grows throughout the range of
strains, even after the stress is nearly stable, and
(3) fabric change lags stress change at small strains, as the small
increase in surface anisotropy contrasts with
substantial increases in deviator stress.
%
%
\subsection{Contacts}\label{sec:contacts}
The mechanical behavior of granular materials is largely determined
by the arrangement and orientations of inter-particle contacts. 
The Satake fabric tensor $\overline{\mathbf{F}}^{\,\text{c}}$ is a measure
of the average contact (``c'') orientation within a granular medium 
\cite{Satake:1982a,Oda:1985a}:
\begin{equation}
  \overline{\mathbf{F}}^{\,\text{c}}=
  \left\langle n_{j}^{\text{c},i}n_{k}^{\text{c},i}\right\rangle
  \mathbf{e}_{j}\otimes\mathbf{e}_{k}
  \label{eq:Fc}
\end{equation}
where $\mathbf{n}^{\text{c},i}$ is the unit normal vector of a single
``$i\,$th'' contact, which is averaged for all contacts within the
medium. 
Because force transmission takes place through
the contacts, this tensor is commonly associated with
stiffness and strength.
Radjai \cite{Radjai:1998a} found that
deviatoric stress is carried primarily through those contacts that
bear a larger-than-mean normal force (also \cite{Thornton:2010a,Guo:2013a}).
This observation has led to a variation of the Satake tensor, by averaging
the contact orientations among this subset of ``strong'' contacts:
\begin{equation}
  \overline{\mathbf{F}}^{\,\text{c-strong}}=
  \left\langle n_{j}^{\text{c-strong},i}n_{k}^{\text{c-strong},i}\right\rangle 
  \mathbf{e}_{j}\otimes\mathbf{e}_{k}
  \label{eq:Fcstrong}
\end{equation}
which has been found to correlate with the deviatoric stress tensor
\cite{Antony:2004b,Thornton:2010a}. 
\par
Another tensor $\overline{\mathbf{G}}^{\,\text{c}}$ associated with
contacting particles is the averaged product of the branch vectors
$\mathbf{l}^{\text{c},i}$ that connect the centers of contacting
particles (e.g., \cite{Pena:2007a}):
\begin{equation}
  \overline{\mathbf{G}}^{\,\text{c}}=
  \frac{1}{\left(D_{50}\right)^{2}}
  \left\langle l_{j}^{\text{c},i}l_{k}^{\text{c},i}\right\rangle
  \mathbf{e}_{j}\otimes\mathbf{e}_{k}
  \label{eq:Lc}
\end{equation}
Magoariec et al. \cite{Magoariec:2008a} suggested this fabric measure
as a possible internal variable for predicting the stress of 2D assemblies
of ellipses. 
The measure reflects both orientation and distance between
the particle pairs and admits possible correlations between the
orientation and distance
\cite{Guo:2013a}. 
The measure has been normalized so that an assembly
of equal-size spheres yields a trace of 1.0. A tensor of strong contacts
$\overline{\mathbf{G}}^{\,\text{c-strong}}$ can also be computed,
in the manner of Eq.~(\ref{eq:Fcstrong}). 
\par
Stress in a granular medium is the volume-average of dyadic products
of branch vectors and contact forces $\mathbf{f}^{\text{c},i}$ among
all contacts within an assembly:
\begin{equation}
  \boldsymbol{\sigma}=\frac{1}{V}
  \sum\mathbf{l}^{\text{c},i}\otimes\mathbf{f}^{\text{c},i}
\end{equation}
Because differences in the orientations of the contact forces
and the contact normals $\mathbf{n}^{\text{c},i}$ are limited by the
friction coefficient, the stress is likely related to the average
of the dyadic products $\mathbf{l}^{\text{c},i}\otimes\mathbf{n}^{\text{c},i}$.
This observation suggests a third, mixed measure of contact orientation:
\begin{equation}
  \overline{\mathbf{H}}^{\,\text{c}}=
  \frac{1}{D_{50}}
  \left\langle l_{j}^{\text{c},i}n_{k}^{\text{c},i}\right\rangle
  \mathbf{e}_{j}\otimes\mathbf{e}_{k}
  \label{eq:Hbarc}
\end{equation}
along with its strong-contact 
counterpart $\overline{\mathbf{H}}^{\,\text{c-strong}}$.
\par
\begin{figure}
  \centering
  \mbox{%
    \subfloat[Anisotropy of $\overline{\mathbf{F}}^{\,\text{c}}$]
    {\includegraphics[width=0.485\textwidth]{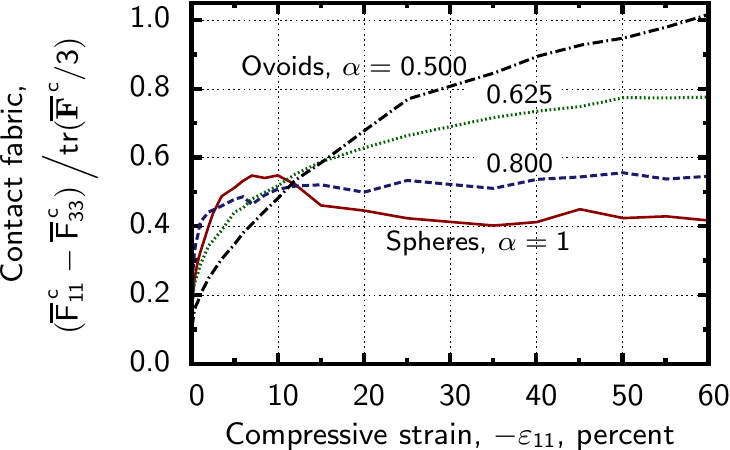}}
    \quad
    \subfloat[Anisotropy of $\overline{\mathbf{H}}^{\,\text{c}}$]
    {\includegraphics[width=0.485\textwidth]{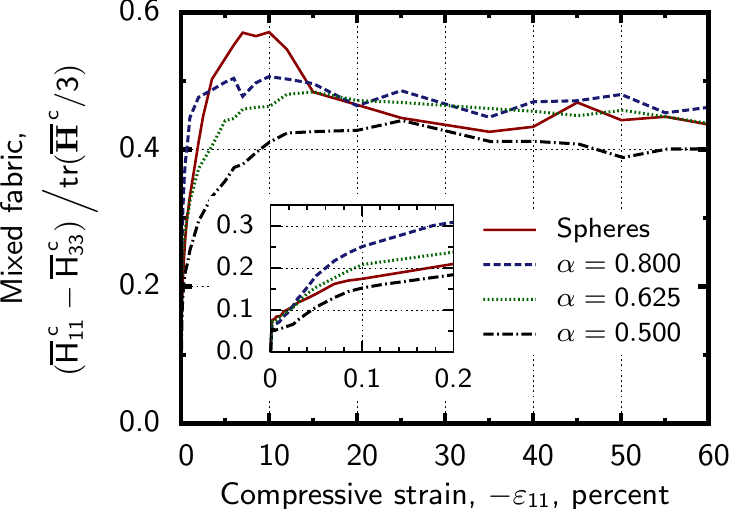}}
  }\\[1ex]
  \mbox{%
    \subfloat[Anisotropy of $\overline{\mathbf{H}}^{\,\text{c-strong}}$]
    {\includegraphics[width=0.485\textwidth]{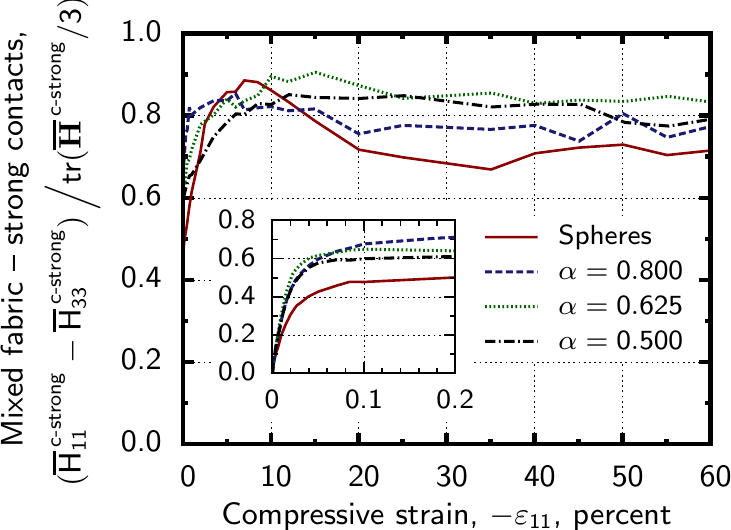}}
    \quad
    \subfloat[Intermediate anisotropy of 
              $\overline{\mathbf{H}}^{\,\text{c-strong}}$]
    {\includegraphics[width=0.485\textwidth]{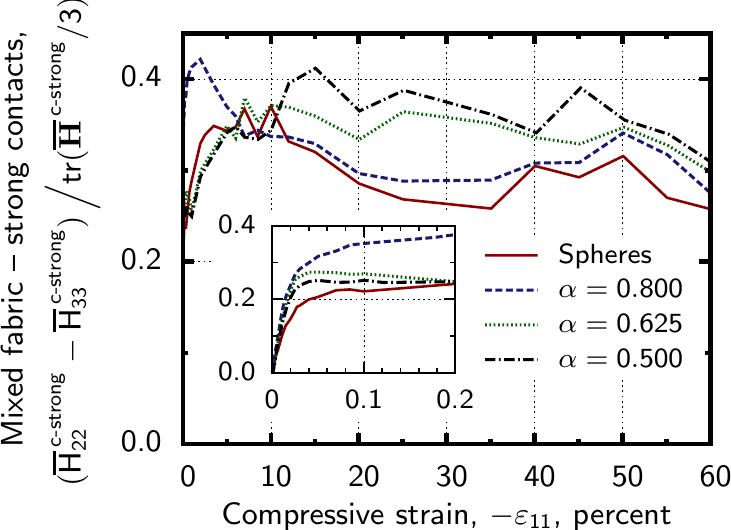}}
  }
  \caption{Evolution of contact anisotropies 
           $\overline{\mathbf{F}}^{\,\text{c}}$,
           $\overline{\mathbf{H}}^{\,\text{c}}$,
           and $\overline{\mathbf{H}}^{\,\text{c-strong}}$ (Eqs.~\ref{eq:Fc}
         and~\ref{eq:Hbarc}) expressed as the differences of their major
         and minor principal values.
         Inset plots detail the small-strain behavior.
         \label{fig:Fcbar_Fcstrongbar}}
\end{figure}
\begin{figure}
  \centering
  \mbox{%
    \subfloat[]{\includegraphics{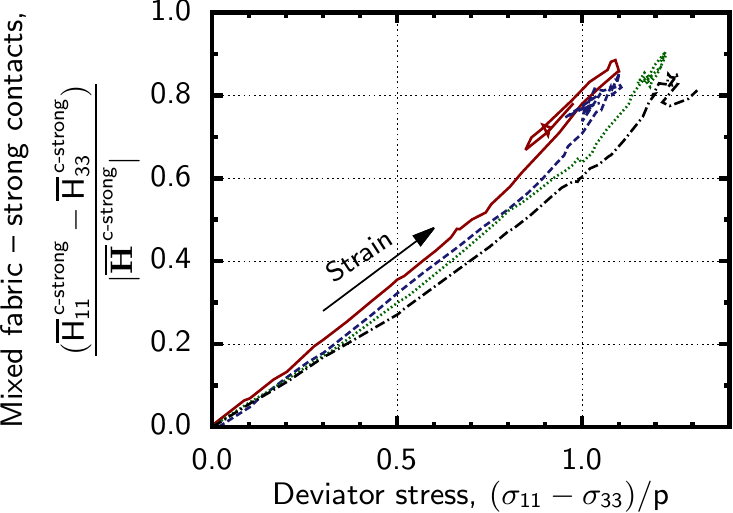}}
    \quad\quad
    \subfloat[]{\includegraphics{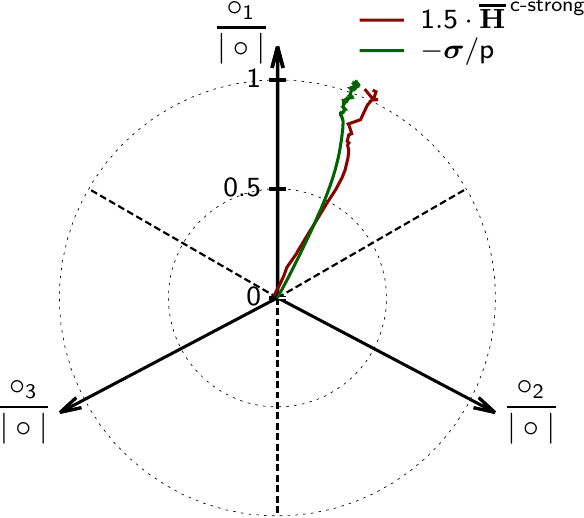}}
  }
  \caption{Correspondence between stress and the mixed-vector contact tensor
           $\overline{\mathbf{H}}^{\,\text{c-strong}}$ 
           during biaxial compression:
           (a) deviatoric stress vs. 
           $\overline{\mathbf{H}}^{\,\text{c-strong}}$ 
           for four particle shapes during strains of 0 to 60\%, 
           (b) $\pi$-plane paths of stress 
           and $\overline{\mathbf{H}}^{\,\text{c-strong}}$ for ovoids
           ($\alpha=0.500$), with the latter tensor scaled by 
           factor 1.5.\label{fig:Corr-Fbarstrongc}}
\end{figure}

These six measures of contact orientation were investigated with the
DEM simulations,
with the intent of determining a fabric measure that
is most closely associated with deviatoric stress.
Some of the results
are illustrated in 
Figs.~\ref{fig:Fcbar_Fcstrongbar} and~\ref{fig:Corr-Fbarstrongc},
and all measures are summarized in Table~\ref{tab:Correlations_FGH}.
Figures~\ref{fig:Fcbar_Fcstrongbar}a,
\ref{fig:Fcbar_Fcstrongbar}b,
and~~\ref{fig:Fcbar_Fcstrongbar}c show the progressions of the
normalized 
deviatoric anisotropies of $\overline{\mathbf{F}}^{\,\text{c}}$,
$\overline{\mathbf{H}}^{\,\text{c}}$, and
$\overline{\mathbf{H}}^{\,\text{c-strong}}$
across the $x_{1}$--$x_{2}$ directions
(for example, in Fig.~\ref{fig:J_bar_p} we plot the difference
$\overline{F}_{11}^{\,\text{c}}-\overline{F}_{33}^{\,\text{c}}$ divided
by the mean 
$|\overline{\mathbf{F}}^{\,\text{c}}|=\text{tr}(\overline{\mathbf{F}}^{\,\text{c}})/3$).
As has been widely reported, contact normals become predominantly
oriented in the direction of compression, with anisotropy increasing
with strain \cite{Oda:1972c,Oda:1985a,Rothenburg:1989a,Ouadfel:2001a,Guo:2013a}.
Kruyt \cite{Kruyt:2012a} and Oda et al. \cite{Oda:1985a} have shown that the 
anisotropy of $\overline{\mathbf{F}}^{\,\text{c}}$
at small strains is primarily the result of contacts being disengaged
in the extension direction (also \cite{Rothenburg:1992b}). 
At larger
strains, changes in $\overline{\mathbf{F}}^{\,\text{c}}$ are also
produced by the reorientation of existing contacts
\cite{Kuhn:2010a}. 
Anisotropy in
contact orientation is larger for the less spherical shapes across
the full range of strains
(see Fig.~\ref{fig:Fcbar_Fcstrongbar}a) \cite{Azema:2007a}.
For the sphere assemblies,
this anisotropy reaches a peak value at 8--10\% strain, which corresponds
to the peak in stress ratio (Fig.~\ref{fig:Stress_strain}a). At
strains beyond 30\%, the sphere assemblies reach the critical state,
in which stress, volume, and contact fabric are stationary, a condition
that is also seen in biaxial loading simulations of disks and spheres
\cite{Pena:2007a,Kruyt:2012a,Zhao:2013a}.
With non-spherical particles, more prolonged deformation is
required to reach a steady fabric
(see Fig.~\ref{fig:Fcbar_Fcstrongbar}a),
more evidence of significant
fabric rearrangements at large strains and an indication that the
steady-state of fabric is attained at strains greater than 60\%.
For all assemblies at small
strains, the rise in the anisotropy among strong contacts, 
for example
$\overline{H}_{11}^{\,\text{c-strong}}-\overline{H}_{33}^{\,\text{c-strong}}$
(Fig.~\ref{fig:Fcbar_Fcstrongbar}c),
occurs more steeply than that of all contacts,
for example
$\overline{H}_{11}^{\,\text{c}}-\overline{H}_{33}^{\,\text{c}}$
(Fig.~\ref{fig:Fcbar_Fcstrongbar}b),
In this regard, the strong-contact measures of fabric more closely
follow the rise in the stress ratio than do measures
that include all contacts.
\par
For sphere assemblies, the deviatoric part of the fabric
tensor $\overline{\mathbf{F}}^{\,\text{c-strong}}$
correlates closely with deviatoric stress, 
a trend noted in \cite{Thornton:2010a,Guo:2013a}.
This trend was not observed with $\overline{\mathbf{F}}^{\,\text{c-strong}}$
for the non-spherical particles, so we searched for a closer
stress-fabric correspondence among the other measures of contact fabric.
In Table~\ref{tab:Correlations_FGH}, we rank the correlations between
deviatoric stress and the deviatoric parts of six contact tensors
with respect to their differences 
$\circ_{11}-\circ_{33}$ and $\circ_{22}-\circ_{33}$
across the full range of strains (0 to 60\%) and for all four
particle shapes.
\begin{table}
  \caption{Average correlations between six contact fabric measures
           and the deviatoric stress
           during biaxial plane strain compression (see Eq.~\ref{eq:c2}).
           \label{tab:Correlations_FGH}}
  \centering%
  \begin{tabular}{lcc}
  \hline
    & \multicolumn{2}{c}{Fabric--stress correlations}\\
    \cline{2-3}
    Fabric tensor, $\circ$ &
    $c_{1}(\circ,\boldsymbol{\sigma})$ &
    $c_{2}(\circ,\boldsymbol{\sigma})$ \\
  \hline 
  $\overline{\mathbf{H}}^{\,\text{c-strong}}$ & 0.993 & 0.988 \\
  $\overline{\mathbf{F}}^{\,\text{c-strong}}$ & 0.941 & 0.985 \\
  $\overline{\mathbf{G}}^{\,\text{c-strong}}$ & 0.843 & 0.979 \\
  $\overline{\mathbf{H}}^{\,\text{c}}$ & 0.875 & 0.840 \\
  $\overline{\mathbf{F}}^{\,\text{c}}$ & 0.800 & 0.810 \\
  $\overline{\mathbf{G}}^{\,\text{c}}$ & 0.640 & 0.842 \\
  \hline 
  \end{tabular}
\end{table}
Correlation is measured with Pearson coefficients ``$c_{1}$''
and ``$c_{2}$'', for example
\begin{equation}
  c_{2}(\overline{\mathbf{H}}^{\,\text{c-strong}},\boldsymbol{\sigma})=
  \frac{\text{cov}
        ( \overline{H}_{22}^{\,\text{c-strong}}
         -\overline{H}_{33}^{\,\text{c-strong}},\:
         \sigma_{22}-\sigma_{33})}
       {\text{std}
        (\overline{H}_{22}^{\,\text{c-strong}}
        -\overline{H}_{33}^{\,\text{c-strong}})
        \:\text{std}(\sigma_{22}-\sigma_{33})}\label{eq:c2}
\end{equation}
with the covariance and standard deviations measured across the full
range of strains for each particle shape.
The complementary
correlation $c_{1}$ applies to differences
$\circ_{11}-\circ_{33}$.
Both correlations are shown in the table. 
Of the six contact orientation tensors,
the mixed-vector orientation $\overline{\mathbf{H}}^{\,\text{c-strong}}$
is the most closely correlated with the stress tensor $\boldsymbol{\sigma}$.
Although $\overline{\mathbf{F}}^{\,\text{c-strong}}$ correlates favorably
as small stresses, 
the correlation is less favorable at stresses beyond 2\%
and for non-spherical shapes. 
The close relationship between $\overline{\mathbf{H}}^{\,\text{c-strong}}$
and stress is shown in Fig.~\ref{fig:Corr-Fbarstrongc}, in which
Fig.~\ref{fig:Corr-Fbarstrongc}a shows the correspondence of
$\overline{H}_{11}^{\,\text{c-strong}}-
\overline{H}_{33}^{\,\text{c-strong}}$ and
$\sigma_{11}-\sigma_{33}$ for the four particle shapes. 
Although the slope in the figure increases with increasing sphericity of the
particles, the relationship for each shape is nearly linear, and even the
brief relapses in stress that occur at large strains
are accompanied by corresponding decreases
in this fabric measure. 
Figure~\ref{fig:Corr-Fbarstrongc}b shows the
evolution of stress and of $\overline{\mathbf{H}}^{\,\text{c-strong}}$
within the $\pi$-plane for ovoids with $\alpha=0.500$. 
By scaling
the path of $\overline{\mathbf{H}}^{\,\text{c-strong}}$ by
a factor of 1.5, the figure reveals a close alignment of the intermediate
principal values of stress and 
those of $\overline{\mathbf{H}}^{\,\text{c-strong}}$.
\subsection{Void space}\label{sec:voids}
Anisotropy of the void space is known to affect the hydraulic properties
of granular, porous materials. 
The void space can be characterized by size,
shape, and connectivity, which can be determined from digitized images
or from the geometric descriptions of particles that are immersed
in the void space.
Experimental techniques, such as X-ray computed tomography (CT)
and digital image correlation (DIC), can be used to track the
evolution of the grain and void spaces,
although the implementation of these methods
is far from trivial
(for example, \cite{Ando:2013a}).
In this work, we quantify the size and shape of
the void space with digitized images
extracted from DEM data; whereas void connectivity and
Minkowski fabric tensors are directly computed from the DEM geometric data.
Together, the two methods are used to characterize the void
fabric with a set of scalar, distribution, matrix, and tensor measures
(see Table~\ref{tab:Measures}, ``Voids, v'').
\par
\begin{figure}
  \centering
  \includegraphics{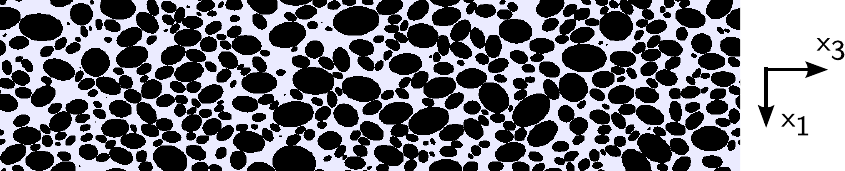}
  \caption{Digitized cross-section through an ovoid assembly 
           ($\alpha=0.625$) at strain
           $\varepsilon_{11}=-0.60$.
           Note that the initial assembly (Fig.~\ref{fig:Assembly})
           was 1.5 times taller (in the $x_{1}$ direction) than
           its width ($x_{2}$ and $x_{3}$ directions), but the assembly
           was greatly squashed and broadened by the vertical
           loading.
           \label{fig:Digitized}}
\end{figure}

The processing of digital images can be performed with discrete
morphological methods \cite{Serra:1982a}, such as dilation, erosion,
opening, and closing. With these methods, an image $X^{\text{v}}$
is represented as an array of 1's and 0's for void and solid space
voxels.
We digitized the DEM assemblies at several strains during
biaxial compression loading. 
The digital density was such that an
average particle was covered by a 20$\times$20$\times$20 grid. 
Figure~\ref{fig:Digitized}
shows a digital image of an $x_{1}$-$x_{3}$ plane through an ovoid
assembly with $\alpha=0.625$.
The periodic boundaries are clearly seen along opposite edges of
the image.
With this image,
the $\varepsilon_{11}$ engineering strain is $-0.60$,
and dilation has increased the $x_{3}$ dimension to 2.67 times its
original width. 
The original assembly, which was 1.5 times taller in
the $x_{1}$ direction (see Fig.~\ref{fig:Assembly}), now has a width
ratio $x_{1}/x_{3}$ of only $0.22$. 
The entire 3D image contains about
60 million voxels.
\par
Hilpert \cite{Hilpert:2003a} proposed a method for estimating the
cumulative distribution $f^{\text{v}}$ of pore size $\rho$ from
3D digital images (see also \cite{Vogel:1997a}):
\begin{equation}
  f_{\rho}^{\text{v}}(\rho)=
  \frac{\text{Vol}\left(\mathcal{O}_{\rho}(X^{\text{v}})\right)}
  {\text{Vol}(X^{\text{v}})}
  \label{eq:fr}
\end{equation}
The void volume in the denominator is a simple counting of the number
of void voxels.
Quantity $\mathcal{O}_{\rho}(X^{\text{v}})$ in the numerator
is a counting of the morphological opening of $X^{\text{v}}$ with a
sphere-shaped
structural template of 1's having radius $\rho$ \cite{Serra:1982a}. 
If the template is a single null
voxel (representing radius 0), the opening operation (i.e., an erosion
followed by a dilation) leaves the image unchanged, and the quotient
is 1.0: 100\% of the void space is larger than size 0. 
When the template
is a digitized ball of radius $\rho$, the quotient is the fraction
of void voxels at a distance greater than $\rho$ from the nearest
particle. 
\par
\begin{figure}
\centering
\includegraphics{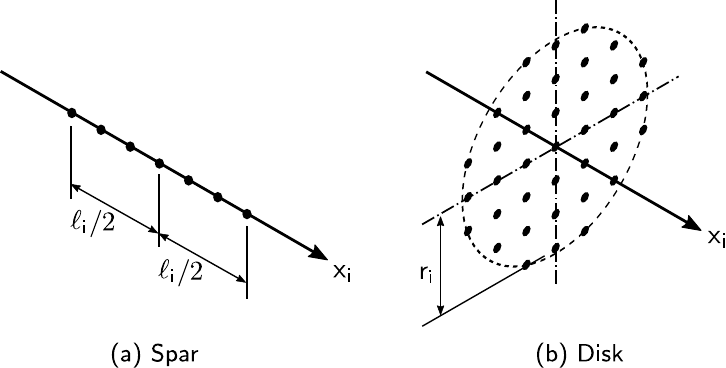}
\caption{Structural templates for characterizing void orientation.
         \label{fig:templates}}
\end{figure}

We generalize the
method by applying two other structural templates (Fig.~\ref{fig:templates}).
To capture the elongation and direction of the void
space, we use a ``spar'' of length $\ell_{i}$ oriented in direction
$x_{i}$: simply a single row of 1's of length $\ell$
along dimension $i$ (Fig.~\ref{fig:templates}a).
This approach yields three
cumulative distributions $f_{l_{i}}^{\,\text{v}}$
for the uninterrupted ``lengths'' $l_{i}$ 
of voids in the three directions, $i=1,2,3$:
\begin{equation}
  f_{\ell_{i}}^{\text{v}}(\ell_{i})=
  \frac{\text{Vol}\left(\mathcal{O}_{\ell_{i}}(X^{\text{v}})\right)}
       {\text{Vol}(X^{\text{v}})}
  \label{eq:fli}
\end{equation}
which represents the fraction of void voxels at a distance greater
than $\ell_{i}/2$ from the nearest particle surface, as measured
in the $x_{i}$ direction. 
This distribution is related to
the mean free path tensor described by Kuo et al. \cite{Kuo:1998a},
which
characterizes the mean separation between particle surfaces
from within the void space.
They approximated this separation $\lambda$
as a function of the measuring direction $\mathbf{n}$:
\begin{equation}
  \lambda(\mathbf{n})\approx
  \overline{\lambda}\left(1+\lambda_{ij}n_{i}n_{j}\right)
\end{equation}
where $\lambda_{ij}$ is the mean free path tensor and $\overline{\lambda}$
is the average separation for all directions and for all points within
the void space. 
Because the median values of the three lengths $l_{i}$
in Eq.~(\ref{eq:fli}) represent the median free paths in directions
$x_{1}$, $x_{2}$, and $x_{3}$, 
a matrix $\overline{\mathbf{L}}^{\,\text{v}}$
can be constructed from these lengths, with the diagonal elements
\begin{equation}
  \overline{L}_{ii}^{\,\text{v}}=
  \frac{3}{\overline{\ell}}\overline{\ell}_{i}
  \label{eq:Lambdaii}
\end{equation}
where $\overline{\ell}_{i}$ is the median value of $\ell_{i}$ for
which $f_{l_{i}}^{\text{v}}(\ell_{i})=0.50$, and 
$\overline{\ell}$
is the trace $\overline{\ell}_{1}+\overline{\ell}_{2}+\overline{\ell}_{3}$.
Our neglect
of off-diagonal terms in Eq.~(\ref{eq:Lambdaii}) assumes an orthotropic
fabric symmetry aligned in the three coordinate directions.
\par
As a further measure of void orientation, we apply a third
structural template to the digitized images:
a disk of pixels having radius $r_{i}$ with an axis of revolution in
the $x_{i}$ direction (Fig.~\ref{fig:templates}b).
This template is used to compute a void distribution
of radial ``breadths''
transverse to the three coordinate directions,
\begin{equation}
  f_{r_{i}}^{\text{v}}(r_{i})=
  \frac{\text{Vol}\left(\mathcal{O}_{r_{i}}(X^{\text{v}})\right)}
       {\text{Vol}(X^{\text{v}})}
  \label{eq:fri}
\end{equation}
and the median values $\overline{r}_{i}$ yield a matrix $\overline{\mathbf{R}}^{\text{v}}$
with diagonal elements
\begin{equation}
  \overline{R}_{ii}^{\,\text{v}}=
  \frac{3}{\overline{r}}\overline{r}_{i}
  \label{eq:Rii}
\end{equation}
which can be thought to represent median directional hydraulic
radii.
The denominator $\overline{r}$ is the trace
$\overline{r}_{1}+\overline{r}_{2}+\overline{r}_{3}$
\par
The connectivity of a 3D void space $X^{\text{v}}$ can be quantified
with the Euler-Poincar\'{e} characteristic $\chi^{\text{v}}(X^{\text{v}})$
\cite{Michielsen:2001a}:
\begin{equation}
  \chi^{\text{v}}(X^{\text{v}})=
  (  \text{no. of connected regions, }K)
   +(\text{no. of cavities, }C)-(\text{no. of tunnels, }G)
  \label{eq:chiv1}
\end{equation}
where ``no.'' means ``number''
If the void space is entirely inter-connected (i.e., with no isolated
void ``bubbles'' inside the solid particles), the number of connected regions
$K$ is one, which is the case with our DEM assemblies. 
The cavities
$C$ within the inter-connected void space are isolated particles
or particle clusters that are disconnected from other particles and
surrounded by void space. 
Because gravity will seat each particle
against other particles, $C$ is one for sands (i.e., a single connected
particle network within the void space $\chi^{\text{v}}$. 
With DEM
simulations that proscribe gravity, however, unconnected ``rattler''
particles can be present and numerous. 
The number of tunnels in a
3D region (that is, the genus $G(X)$ of the region)
is a topological quantity: the
maximum number of full cuts that can be made without producing more
separated (void) regions. 
The genus $G$ can be derived by constructing
the void connectivity graph in which pore bodies (represented as graph
nodes) are connected through restricted passageways (pore throats,
represented as graph edges) between particles 
\cite{DeHoff:1972a,Kwiecien:1990a,Reeves:1996a,Hilpert:2003a}.
This full void graph can be represented as the reduced medial axis
(skeleton or deformation retract) of the void space 
\cite{Serra:1982a,Adler:1992a,Lindquist:1996a,Liang:2000a}.
Genus $G$ (in Eq.~\ref{eq:chiv1})
of the void space is \cite{Adler:1992a}
\begin{equation}
  G(X^{\text{v}})=
  1+(\text{no. of pore throats})-(\text{no. of pore bodies})
  \label{eq:Genus}
\end{equation}
When $K=1$ and $C=1$, a large positive genus $G$ in Eq.~(\ref{eq:Genus})
(or a large negative value of $\chi^{\text{v}}$ in Eq.~\ref{eq:chiv1})
indicates many redundant pathways (i.e., pore throats or tunnels) for
fluid migration through the void space. Prasad et al. \cite{Prasad:1991a}
presented a corresponding formula for the genus of the solid phase.
\par
Although the Euler-Poincar\'{e} characteristic of a sand specimen
is usually approximated by performing morphological
operations on digitized
images \cite{Serra:1982a}, DEM simulations permit the direct computation
of $\chi$ by applying Minkowski functionals to the geometric descriptions
of particle shapes. 
Minkowski functionals (i.e. Minkowski scalars) arise in integral geometry
as four independent scalar values, which include volume and surface area,
that are associated with a $3$-dimensional (3D)
geometric object and are additive and
invariant with respect to translation or rotation of the object. 
The complete set of Minkowski $\nu$-functionals $W_{\nu}$ of a
3D object $X$ are given in the top part of Table~\ref{tab:Minkowski},
adapted from the summary of 
Schr\"{o}der-Turk et al. \cite{SchroderTurk:2011a}.
The table applies to a region $X$ that is a finite union of convex
(but possibly disconnected) objects.
\begin{table}
  \caption{Selected Minkowski functionals and tensors for 
           3D objects $X$, adapted from Schr\"{o}der-Turk et al. 
           \cite{SchroderTurk:2011a}.\label{tab:Minkowski}}
  \centering\small%
  \begin{tabular}{lll}
  \hline 
  Type & Symbol & Definition\\
  \hline 
  Functionals & $W_{0}(X)$ & $\int_{X}dV$\\
   & $W_{1}(X)$ & $\frac{1}{3}\int_{\partial X}dA$\\
   & $W_{2}(X)$ & $\frac{1}{3}\int_{\partial X}\frac{1}{2}\left(\kappa_{1}+\kappa_{2}\right)\, dA$\\
   & $W_{3}(X)$ & $\frac{1}{3}\int_{\partial X}\kappa_{1}\kappa_{2}\, dA$\\
  \hline 
  Tensors & $\mathbf{W}_{1}^{2,0}(X)$ & $\frac{1}{3}\int_{\partial X}\mathbf{x}\otimes\mathbf{x}\, dA$\\
   & $\mathbf{W}_{1}^{0,2}(X)$ & $\frac{1}{3}\int_{\partial X}\mathbf{n}\otimes\mathbf{n}\, dA$\\
   & $\mathbf{W}_{3}^{2,0}(X)$ & $\frac{1}{3}\int_{\partial X}\kappa_{1}\kappa_{2}\mathbf{x}\otimes\mathbf{x}\, dA$\\
  \hline 
  \end{tabular}
\end{table}
In the expressions, $\kappa_{1}$and $\kappa_{2}$ are the principal
curvatures of the object's surface $\partial X$,
quantity $(\kappa_{1}+\kappa_{2})/2$
is the mean curvature, and $\kappa_{1}\kappa_{2}$ is the Gaussian
curvature. Functional $W_{0}$ is the volume; $W_{1}$ is one-third
of the surface area; $W_{2}$ is equal to $2\pi/3$ times the ``mean
breadth'' $B(X)$ of the object; and $W_{3}$ is directly related
to the characteristic $\chi$ (see Eq.~\ref{eq:chiv1}) as
\begin{equation}
W_{3}=\frac{4\pi}{3}\chi(X)
\end{equation}
which is a form of the Gauss-Bonnet formula. 
Evaluating functionals $W_{2}$
and $W_{3}$ for shapes with sharp edges or corners requires cylindrical
or spherical rounding (creating a smooth, differentiable surface)
and then finding the integral limit as the radius is reduced to zero.
\par
Functional $W_{3}$ is $4\pi/3$ for a solid ball, 
and $W_{3}$ is $8\pi/3$
for two disjoint balls ($\chi=1$ and $2$ respectively). If two balls
are brought into contact, forming a finite contact area, $W_{3}$
is reduced from $8\pi/3$ to $4\pi/3$: the two spherical surfaces
have a positive Gaussian curvature $\kappa_{1}\kappa_{2}$ and together
contribute $8\pi/3$ to the integral, but the bridge between the two
spheres has a negative curvature and contributes $-4\pi/3$. 
By extension,
the Euler-Poincar\'{e} characteristic $\chi^{\text{s}}$ of an assembly
of connected solid particles $X^{\text{s}}$ is
\begin{equation}
  \chi^{\text{s}}(X^{\text{s}})=
  \frac{3}{4\pi}W_{3}=
  1+(\text{no. of particles})-(\text{no. of contacts between particles})
  \label{eq:Chis1}
\end{equation}
Because the void space and solid space share the same surface $\partial X$
with the same Gaussian curvature, the Euler-Poincar\'{e}
characteristic of the void space is also 
\begin{equation}
  \chi^{\text{v}}=\chi^{\text{s}}
  \label{eq:Chiv2}
\end{equation}
This approach to quantifying $\chi$ for a DEM assembly (or its void
space) involves simply counting the numbers of particles and contacts
(as in Eq.~\ref{eq:Chis1})
and does not require the direct evaluation of surface integrals. 
The void connectivity is normalized as
\begin{equation}
  \overline{\chi}^{\text{v}}=\chi^{\text{v}}/N^{\text{p}}
  \label{eq:Chivbar}
\end{equation}
by dividing by the number of particles $N^{\text{p}}$. 
Note that
$\chi^{\text{v}}$ in Eq.~(\ref{eq:Chis1}) is
directly related to the degree of structural redundancy of the particle
network \cite{Thornton:1998a,Kruyt:2009a}.
\par
\begin{figure}
  \centering
  \includegraphics{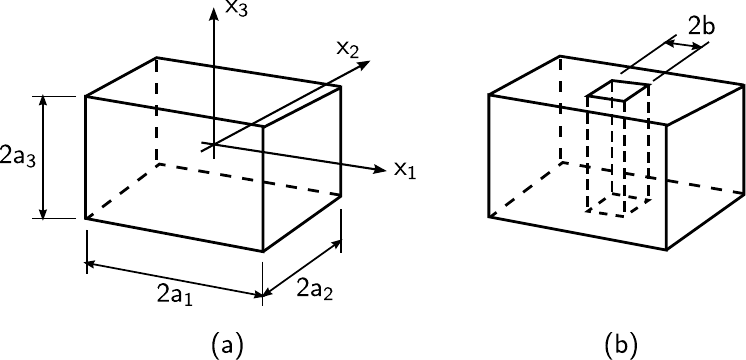}
  \caption{Rectangular blocks used in an example
           of Minkowski tensors.\label{fig:Blocks}}
\end{figure}

Beyond Minkowski functionals,
Minkowski \emph{tensors} provide measures of the shape and orientation of
the void space. 
Schr\"{o}der-Turk et al. \cite{SchroderTurk:2010a,SchroderTurk:2011a}
identify six rank-two Minkowski tensors that form a complete set
of isometry covariant, additive and continuous functions for three
dimensional poly-convex bodies. 
Three of the six tensors are given in Table~\ref{tab:Minkowski}. 
The first two, $\mathbf{W}_{1}^{2,0}$
and $\mathbf{W}_{1}^{0,2}$, have been applied in the definitions
of the surface inertia and surface normal tensors: 
the $\overline{\mathbf{I}}^{\text{s}}$
and $\overline{\mathbf{S}}^{\text{s}}$ of Eqs.~(\ref{eq:Ibari}),
(\ref{eq:Ipar}), and~(\ref{eq:Sbar}). 
The final tensor is covariant
with respect to translation and rotation; depends solely on the shape,
size, and connectivity of a 3D object; and can be directly evaluated
from DEM geometric data for either the solid or void regions:
\begin{equation}
  \mathbf{W}_{3}^{2,0}(X)=
  \frac{1}{3}\int_{\partial X}\kappa_{1}\kappa_{2}\mathbf{x}
  \otimes\mathbf{x}\, \DIF A
  \label{eq:W202}
\end{equation}
The meaning of this tensor
(and the corresponding functional $W_{3}$) is illustrated
with Fig.~\ref{fig:Blocks} for the cases of a rectangular block
of size $2a_{1}\times2a_{2}\times2a_{3}$ and of the block pierced
by a square tunnel of size $2a_{1}\times2b\times2b$. .
The sides and edges of the block have zero Gaussian curvature and
do not contribute to $W_{3}$ or to $\mathbf{W}_{3}^{2,0}$, whose values
are derived entirely from the corners. 
The contribution to $W_{3}$
of a single corner is $\gamma/3$, where $\gamma$ is the 
Descartes angular
deficit of the corner ($\pi/2$ for a square corner). 
The value of
$W_{3}$ for the eight corners of a solid block is $8(\pi/2)/3=4\pi/3$.
If the coordinate system is centered within the block, 
tensor $\mathbf{W}_{3}^{2,0}$
is formed from the dyads $\mathbf{x}\otimes\mathbf{x}$, where vectors
$\mathbf{x}$ are directed from the center to the corners:
\begin{equation}
  \mathbf{W}_{3}^{2,0}
  (X\text{ of Fig. \ref{fig:Blocks}a})=\frac{1}{3}8\frac{\pi}{2}
  \left[\begin{array}{ccc}
        a_{1}^{2} & 0 & 0\\
        0 & a_{2}^{2} & 0\\
        0 & 0 & a_{3}^{2}
       \end{array}\right]
  \label{eq:blocka}
\end{equation}
capturing information of the block's shape in Fig.~\ref{fig:Blocks}a. 
The value of $\mathbf{W}_{3}^{2,0}$
for the pierced block (Fig.~\ref{fig:Blocks}b) equals
the expression (\ref{eq:blocka}) plus contributions
from the eight interior corners. 
Each of these corner has a negative
angular deficit, $-\pi/2$, giving
\begin{equation}
  \mathbf{W}_{3}^{2,0}(X\text{ of Fig. \ref{fig:Blocks}b})=
  \frac{4\pi}{3}\left[\begin{array}{ccc}
                      a_{1}^{2}-b^{2} & 0 & 0\\
                      0 & a_{2}^{2}-b^{2} & 0\\
                      0 & 0 & 0
                      \end{array}\right]
\end{equation}
A tunnel is seen to modestly reduce the first two diagonal terms,
while reducing the third term to zero. 
Multiple tunnels in the $x_{3}$
direction will make the third term negative, an indicator of multiple
pathways (and void anisotropy) in this direction. 
\par
The tensor $\mathbf{W}_{3}^{2,0}$ of a single object $X$ whose (local)
center is offset by vector $\mathbf{t}$ from the origin of a 
(global) coordinate
frame is given by Schr\"{o}der-Turk et al. 
(in \cite{SchroderTurk:2011a}, their Eq.~6):
\begin{equation}
  \mathbf{W}_{3}^{2,0}=
  \breve{\mathbf{W}}_{3}^{2,0}
   +2\cdot\mathbf{t}\otimes
   \left(\frac{1}{3}\int_{\partial X}\kappa_{1}\kappa_{2}
   \mathbf{\breve{x}}\, \DIF A\right)
   +W_{3}\mathbf{t}\otimes\mathbf{t}
  \label{eq:W203all}
\end{equation}
which is the parallel-axis relationship for $\mathbf{W}_{3}^{2,0}$.
In this equation, $\mathbf{\breve{W}}_{3}^{2,0}$ and $\mathbf{\breve{x}}$
are measured relative to the local center of the object, and the expression
in parentheses is the
local Minkowski vector $\breve{\mathbf{W}}_{3}^{1,0}$,
which is equal to zero for spheres, ovoids, and ellipsoids and other objects
having orthorhombic symmetry.
\par
For a granular assembly, region $X$ can represent the solid particles,
which are joined at their contacts. 
Combining equations~(\ref{eq:W202})
and~(\ref{eq:W203all}); noting that $W_{3}=4\pi/3$ for a solid
particle without holes and that $W_{3}=-4\pi/3$ for a contact bridge;
and assuming that $\breve{\mathbf{W}}_{3}^{1,0}=0$
for each particle, we have
\begin{equation}
  \mathbf{W}_{3}^{2,0}=
  \sum_{p=1}^{N^{p}}\left(\breve{\mathbf{W}}_{3}^{2,0,p}
  +\frac{4\pi}{3}\mathbf{x}^{p}\otimes\mathbf{x}^{p}\right)
  -\sum_{c=1}^{N^{c}}\frac{4\pi}{3}\mathbf{x}^{c}\otimes\mathbf{x}^{c}
  \label{eq:W320}
\end{equation}
In this expression,
contributions are summed from the $N^{p}$ particles and the $N^{c}$
contacts: $\breve{\mathbf{W}}_{3}^{2,0,p}$ is the local tensor for
particle $p$, $\mathbf{x}^{p}$ is the vector from the assembly's
center to a particle's center, and $\mathbf{x}^{c}$ is the vector
from the assembly's center to a contact.
From the example in Fig.~\ref{fig:Blocks}, we note that the magnitudes
of the components of tensor $\mathbf{W}_{3}^{2,0}$ depend upon the
overall shape and size of the
region $X$ as well as on connectivity within the
region.
During the simulated
compression of our box-shaped granular assembly, its overall
shape changes from tall to squat 
(compare Figs.~\ref{fig:Assembly} and~\ref{fig:Digitized}).
To compensate for this change in shape,
we divide $\mathbf{W}_{3}^{2,0}$
by the integral in Eq.~(\ref{eq:W202}), as applied to the full 
assembly's boundary:
\begin{equation}
\overline{\mathbf{W}}_{3}^{\,\text{v},2,0}=
(\mathbf{W}_{3}^{2,0})^{-1}_{\text{Eq.~\ref{eq:W202}, boundaries}}
\cdot (\mathbf{W}_{3}^{2,0})_{\text{Eq.~\ref{eq:W320}}}
\label{eq:MinkNormal}
\end{equation}
For the rectangular assembly of our simulations, the boundary
integral is simply that of a rectangular block, as in Eq.~(\ref{eq:blocka}).
Anisotropies in the void shape and connectivity are measured with
this tensor.
%
%
\par
\begin{figure}
  \centering
  \parbox{3.00in}{\centering%
    \subfloat[]{\includegraphics{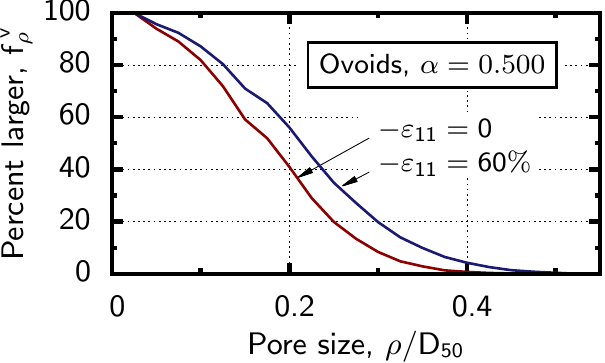}}\\[1ex]
    \subfloat[]{\includegraphics{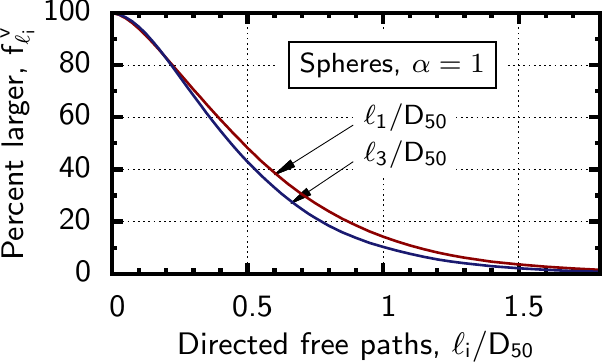}}\\[1ex]
    \subfloat[]{\includegraphics{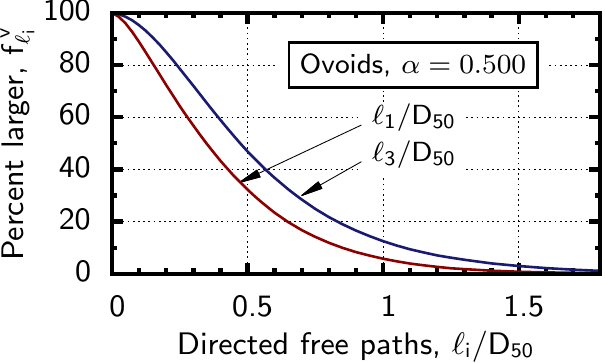}}\\[1ex]
    \subfloat[]{\includegraphics{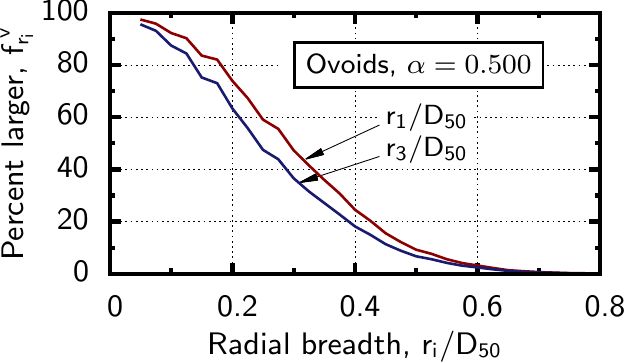}}
  }
  \caption{Distribution of void sizes and orientations: 
  (a) pore size distribution
  $f^{\text{v}}_{\rho}$ for ovoid assemblies ($\alpha=0.500$) 
  at strains of 0\% and 60\% (Eq.~\ref{eq:fr}), 
  (b) directed free path
  distributions $f^{\text{v}}_{\ell_{i}}$
  for sphere assemblies in 
  $x_{1}$ and $x_{3}$ directions at strain 60\% 
  (Eq.~\ref{eq:fli}),
  (c) directed free path
  distributions $f^{\text{v}}_{\ell_{i}}$
  for ovoid assemblies ($\alpha=0.500$)
  at strain 60\% (Eq.~\ref{eq:fli}), and
  directed radial breadth distributions $f^{\text{v}}_{r_{i}}$
  for ovoid assemblies ($\alpha=0.500$)
  at strain 60\% (Eq.~\ref{eq:fri}).
  \label{fig:VoidDistrib}}
\end{figure}

We now apply the various void fabric measures to the
DEM simulations of biaxial compression.
Distributions of void size and orientation are
shown in Fig.~\ref{fig:VoidDistrib}.
The shift in the pore size distributions
between the strains $-\varepsilon_{11}=0\%$
and 60\% (Fig.~\ref{fig:VoidDistrib}a)
is due to material dilation, causing an increase in void volume.
In this figure, void dimensions are normalized by dividing
by the median particle size $D_{50}$.
Distributions of free path distances directed in the $x_{1}$ and
$x_{3}$ directions,
$f^{\text{v}}_{\ell_{1}}(\ell_{1})$ and $f^{\text{v}}_{\ell_{3}}(\ell_{3})$,
are shown in Fig.~\ref{fig:VoidDistrib}b 
for the sphere assembly and in Fig.~\ref{fig:VoidDistrib}c
for the ovoid assembly, both at the final strain of 60\%. 
The void orientation is clearly different for the two types of particles.
For spheres at strain 60\%,
the voids are slightly longer in the $x_{1}$ direction
(i.e., the direction of compressive
loading). 
This observation is consistent with that of Oda and his coworkers
\cite{Oda:1998b}, who found that columnar voids form between
chains of heavily loaded circular disks
and that the columns and chains were oriented predominantly
in the direction of compression. 
For the flattest ovoid particles
($\alpha=0.500$), however, the directed free paths are \emph{shorter} in
the direction of compressive loading. As has been seen, elongated
particles become oriented with long axes more aligned in the direction
of extension (Fig.~\ref{fig:J_bar_p}).
The voids become elongated in this same direction, as
can be discerned in the cross-section of Fig.~\ref{fig:Digitized}
($\alpha=0.625$).
Figure~\ref{fig:VoidDistrib}d shows the corresponding distribution
of the directed radial breadths of the voids for the 
ovoid assembly at the final strain of 60\%.
Comparing Figs.~\ref{fig:VoidDistrib}c
and~\ref{fig:VoidDistrib}d, we see that the voids have become
elongated in the $x_{1}$ direction while become narrower in the
transverse directions.
\par
\begin{figure}
  \centering
  \mbox{%
    \subfloat[Anisotropy across directions $x_{1}$--$x_{3}$]
      {\includegraphics[width=0.485\textwidth]{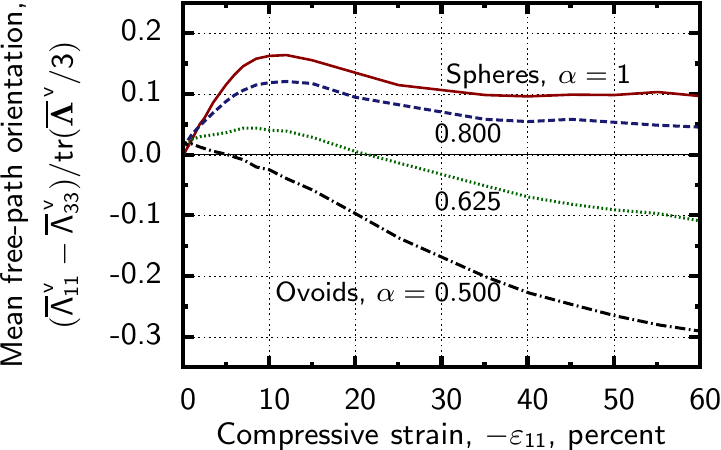}}
      \quad
    \subfloat[Anisotropy across directions $x_{2}$--$x_{3}$]
      {\includegraphics[width=0.485\textwidth]{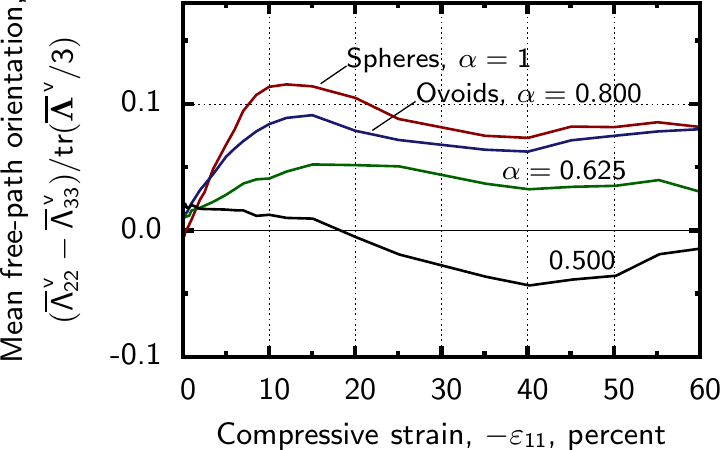}}
  }
  \caption{Anisotropy of median free-path tensor 
           $\overline{\mathbf{L}}^{\,\text{v}}$
           of the voids during biaxial compression.
           \label{fig:Lambda_vs_strain}}
\end{figure}
\begin{figure}
  \centering
  \mbox{%
    \subfloat[Anisotropy across directions $x_{1}$--$x_{3}$]
      {\includegraphics[width=0.485\textwidth]{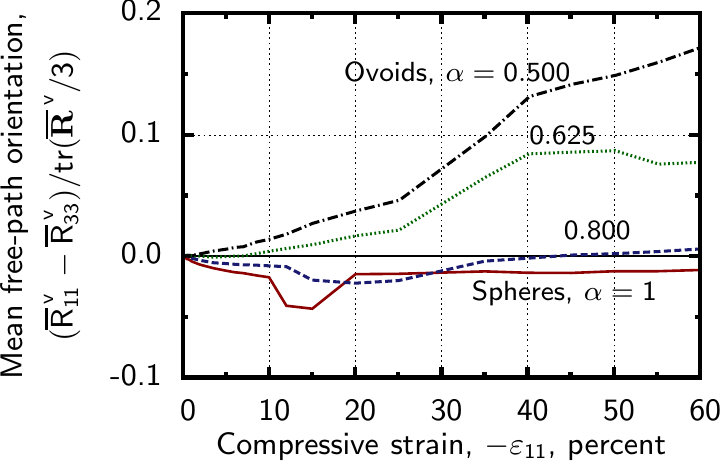}}
      \quad
    \subfloat[Anisotropy across directions $x_{2}$--$x_{3}$]
      {\includegraphics[width=0.485\textwidth]{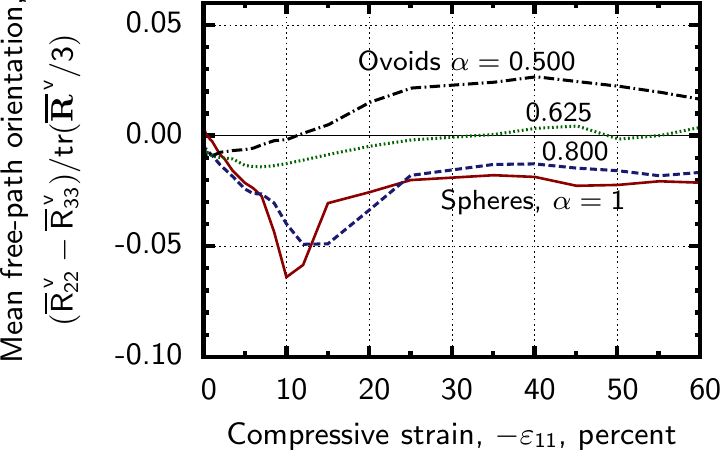}}
  }
  \caption{Anisotropy of median radial breadth tensor 
           $\overline{\mathbf{R}}^{\,\text{v}}$
           of the voids during biaxial compression.
           \label{fig:LLambda_vs_strain}}
\end{figure}

This difference in void shape within
assemblies of spheres and within those of flattened shapes is
also evident in the evolution of the median free path orientation matrix
of Eq.~(\ref{eq:Lambdaii}) for the four particle shapes
(see Fig.~\ref{fig:Lambda_vs_strain}).
Spheres and the most rotund ovoids ($\alpha=0.800$) develop voids
that are longer in the $x_{1}$ compression direction
($\overline{L}_{11}^{\,\text{v}}-\overline{L}_{33}^{\,\text{v}}>0$);
whereas, the flatter ovoids develop voids that are
longer in the extension
direction. 
With these flatter particles, the void elongation continues to change
at strains beyond 60\%.
Figure~\ref{fig:LLambda_vs_strain}
shows the corresponding anisotropy of the median
radial breadth tensor $\overline{\mathbf{R}}^{\text{v}}$.
Comparing the two figures,
\ref{fig:Lambda_vs_strain} and \ref{fig:LLambda_vs_strain},
we see countervailing trends of the two anisotropy measures:
as the voids become more elongated in one direction
(Fig.~\ref{fig:Lambda_vs_strain},
as developed with the structural template
of Fig.~\ref{fig:templates}a), the voids become
narrower in the transverse directions
(Fig.~\ref{fig:LLambda_vs_strain},
as developed with the structural template
of Fig.~\ref{fig:templates}b), so that an increase in the
anisotropy $\overline{\mathbf{L}}^{\,\text{v}}$ is
accompanied by a counter-anisotropy of
$\overline{\mathbf{R}}^{\,\text{v}}$.
Although both $\overline{\mathbf{L}}^{\,\text{v}}$
and $\overline{\mathbf{R}}^{\,\text{v}}$ are extracted from the void space images,
no attempt was made to correlate void ratio with these tensors.
All simulations began dense of the critical state, resulting
in significant dilation for all assemblies, so that the mean
void dimensions (as measured by $f^{\text{v}}_{\rho}(\rho )$ or the
traces of $\overline{\mathbf{L}}^{\,\text{v}}$
and $\overline{\mathbf{R}}^{\,\text{v}}$) increased during loading.
\par
\begin{figure}
  \centering
  \includegraphics{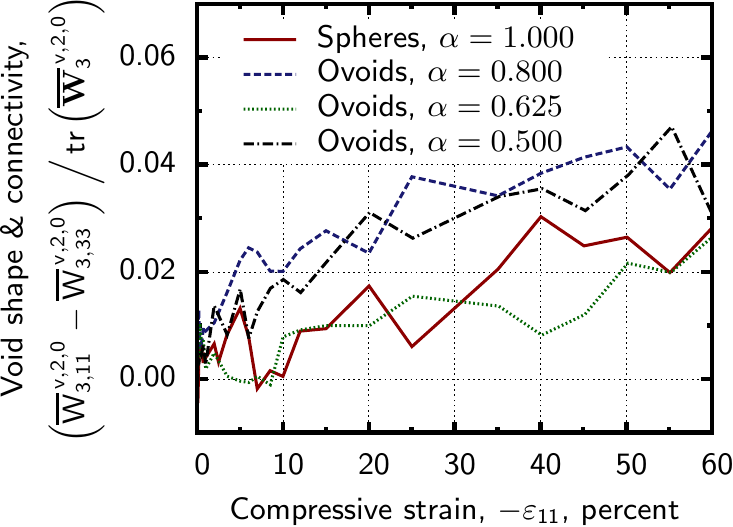}
  \caption{Anisotropy of the Minkowski tensor
  $\overline{\mathbf{W}}_{3}^{2,0}$.\label{fig:Wbar}}
\end{figure}

Void anisotropy is also measured with the normalized Minkowski tensor
$\overline{\mathbf{W}}_{3}^{\text{v},2,0}$ (see Eq.~\ref{eq:MinkNormal}).
Figure~\ref{fig:Wbar} shows the deviator of this tensor divided by its
trace.
The somewhat erratic progression indicates a strong sensitivity of
this measure to subtle changes in particle arrangement, and, as such,
this measure of void anisotropy is not appropriate unless large
numbers of particles can be sampled.
The figure indicates that the anisotropy in
$\overline{\mathbf{W}}_{3}^{\,\text{v},2,0}$
increases during loading, following a similar trend as that
of the measures $\overline{\mathbf{J}}^{\text{p}}$,
$\overline{\mathbf{I}}^{\text{s}}$, and
$\overline{\mathbf{S}}^{\text{s}}$ in
Figs.~\ref{fig:J_bar_p} and~\ref{fig:Surfaces_vs_strain}.
\subsection{Summary of fabric measures}
In this section, we have considered thirteen measures
of fabric anisotropy
and their evolution during biaxial loading.
Three (possibly four) of these measures are closely related to
particle orientation, and any one of the three would
serve as a fabric measure in this regard:
the particle orientation tensor
($\overline{\mathbf{J}}^{\text{p}}$) and
the two tensors of particle surface
orientation ($\overline{\mathbf{I}}^{\text{s}}$ and
$\overline{\mathbf{S}}^{\text{s}}$).
The Minkowski void tensor
$\overline{\mathbf{W}}^{\text{v},2,0}_{3}$ also
follows these same trends.
\par
Six measures of contact orientation were also presented,
involving contact orientation, branch vector
orientation, and mixed contact-branch orientation, and in
which we include either all
contacts or only the strong-contact subset.
Of these measures, the one requiring the
most information, the mixed tensor of strong-contacts
$\overline{\mathbf{H}}^{\text{c-strong}}$,
is most closely correlated with stress evolution.
Particle orientation and surface orientation, however,
are poorly correlated with the deviatoric stress.
\par
Several measures of void orientation were considered: directional
distributions of the free paths
$f^{\text{v}}_{\ell_{i}}(\ell^{i})$
and transverse radial breadths
$f^{\text{v}}_{r_{i}}(r^{i})$,
corresponding orientation matrices
$\overline{\mathbf{L}}^{\text{v}}$ of median free paths
and $\overline{\mathbf{R}}^{\text{v}}$ of radial breadths,
and the average Minkowski tensor
$\overline{\mathbf{W}}_{3}^{\,\text{v},2,0}$.
The evolutions of all of these void measures
(except for $\overline{\mathbf{W}}_{3}^{\,\text{v},2,0}$)
exhibit trends
that are quite different from those of the particle bodies, surfaces,
or contacts, in that opposite trends are found
in their evolution for the least flattened
($\alpha=1$ and 0.800) and most flattened
particles ($\alpha=0.625$ and 0.500, Fig.~\ref{fig:Lambda_vs_strain}).
As for the Minkowski measure
$\overline{\mathbf{W}}_{3}^{\,\text{v},2,0}$,
although its calculation requires more extensive
information and it holds the promise of capturing both the
orientation and the topology of the void space,
it results in a more erratic evolution that follows a
trend much like that of the simpler measures of particle and surface
orientation.
These fabric measures will now be investigated in relation to
the load-induced anisotropies of stiffness and permeability.
\section{Stiffness anisotropy}\label{sec:stiffness}
The strength of granular materials is known to depend on the initial,
deposition anisotropy \cite{Konishi:1982a,Tatsuoka:1990a}, and when
loaded in a particular direction,
strength can also depend upon the anisotropy
that is induced by a previous loading in another direction
\cite{Arthur:1977b,Ishibashi:1991a}.
The relationship between the current stress and the current
fabric has been expressed in so-called stress-force-fabric relations
(e.g., those of \cite{Bathurst:1990a,Ouadfel:2001a,Azema:2007a,Guo:2013a}).
Rather than investigating such effects of anisotropy on the stress
and on the eventual
strength, we focus instead
on the \emph{incremental} stiffness and the influence
of previous loading on this stiffness.
We began with
the same four assemblies, which had a nearly isotropic initial fabric and
were confined with an isotropic stress. 
The monotonic loading in our 
simulations --- biaxial plane strain compression --- induced
an orthotropic
symmetry of the fabric, stress, and stiffness,
with principal directions aligned with the coordinate axes.
\par
\emph{Elastic solids} with
orthotropic symmetry exhibit the following compliance relation:
\begin{equation}
\left[\begin{array}{c}
\DIF\varepsilon_{11}\\
\DIF\varepsilon_{22}\\
\DIF\varepsilon_{33}\\
\DIF\varepsilon_{12}\\
\DIF\varepsilon_{13}\\
\DIF\varepsilon_{23}
\end{array}\right]
=
\left[\begin{array}{cccccc}
   \displaystyle \frac{1}{E_{1}} 
 & \displaystyle \frac{-\nu_{12}}{E_{2}} 
 & \displaystyle \frac{-\nu_{13}}{E_{3}}\\
   \displaystyle \frac{-\nu_{21}}{E_{1}} 
 & \displaystyle \frac{1}{E_{2}} 
 & \displaystyle \frac{-\nu_{23}}{E_{3}}\\
   \displaystyle \frac{-\nu_{31}}{E_{1}} 
 & \displaystyle \frac{-\nu_{32}}{E_{2}} 
 & \displaystyle \frac{1}{E_{3}}\\
 &  &  & \displaystyle \frac{1}{G_{12}}\\
 &  &  &  & \displaystyle \frac{1}{G_{13}}\\
 &  &  &  &  & \displaystyle \frac{1}{G_{23}}
\end{array}\right]
\left[\begin{array}{c}
\DIF\sigma_{11}\\
\DIF\sigma_{22}\\
\DIF\sigma_{33}\\
\DIF\sigma_{12}\\
\DIF\sigma_{13}\\
\DIF\sigma_{23}
\end{array}\right]
\label{eq:Elastic}
\end{equation}
which involves nine material properties 
(note $\nu_{12}=\nu_{21}$, $\nu_{13}=\nu_{31}$, $\nu_{23}=\nu_{32}$). 
With granular materials, however, an initial deviatoric strain of
as small as 0.01\% is sufficient to produce plastic deformation, alter
the incremental stiffness, and disrupt any previous fabric symmetries
\cite{Dobry:1982a}. 
Because of this complexity, which induces non-linearity,
inelasticity, and anisotropy
at the very start of loading, we abandon an assumption of uniform
linearity and leave aside Eq.~(\ref{eq:Elastic}) in the following.
We instead assume a more general behavior that is 
rate-independent and incrementally non-linear
but is positively homogeneous and dependent on the current fabric.
The general compliance and stiffness response operators are
\begin{equation}
 \DIF\boldsymbol{\varepsilon}=\mathbf{f}(\mathcal{S},\DIF\boldsymbol{\sigma}),\quad
 \DIF\boldsymbol{\sigma}=\mathbf{f}^{-1}(\mathcal{S},\DIF\boldsymbol{\epsilon})
\end{equation}
satisfying
\begin{equation}
  \lambda\, \DIF\boldsymbol{\varepsilon}=
  \mathbf{f}(\mathcal{S},\lambda\, \DIF\boldsymbol{\sigma}),
  \quad
  \lambda\, \DIF\boldsymbol{\sigma}=
  \mathbf{f}^{-1}(\mathcal{S},\lambda\, \DIF\boldsymbol{\epsilon})
\end{equation}
for positive scalar $\lambda$ (see Darve~\cite{Darve:1990a}). 
Here $\mathcal{S}$ represents the
current state of the material as characterized by its stress, fabric,
stress history, etc. 
The homogeneous tensor response function $\mathbf{f}$ 
depends upon both the direction
and magnitude of the loading increment $\DIF\boldsymbol{\sigma}$. 
\par
To characterize the behavior at a particular state, we followed a program
suggested by Darve and Roguiez \cite{Darve:1999b}, measuring the
multi-directional incremental stiffnesses at various stages of
biaxial loading. 
We applied both loading and unloading increments in
three directions to DEM assemblies having a current orthotropic symmetry
of fabric
that had been induced by the initial monotonic loading of
biaxial compression.
At various strains during this monotonic
loading, we stopped the simulation and applied the six
incremental oedometric conditions:
\begin{equation}
  \DIF\varepsilon_{ii}\neq0\text{ with }\DIF\varepsilon_{jj}=
  \DIF\varepsilon_{kk}=0,\quad i\neq j\neq k
  \label{eq:Darve}
\end{equation}
in which $i=1,2,3$ and with increments $d\varepsilon_{ii}$ that
were alternatively positive and negative. For example, after biaxial
plane strain compression to a strain of 5\% (i.e., $\varepsilon_{11}=-0.05$),
one incremental loading consisted of a small compressive increment
$\DIF\varepsilon_{22}<0$ while maintaining constant normal strains
$\varepsilon_{11}$ and $\varepsilon_{33}$ (i.e., constant assembly widths). 
As a second loading, a small tensile
increment $\DIF\varepsilon_{22}>0$ was applied, also maintaining constant
$\varepsilon_{11}$ and $\varepsilon_{33}$.
A small increment $\DIF\varepsilon_{ii}=\pm$0.005\% 
was used in our simulations.
No shearing strains (e.g. $\gamma_{12}$,
etc.) were applied in our program, so that each of the six loading
increments maintained the original principal directions of orthotropic
fabric and stress. 
Throughout the program, fabric and stress
changed, but their principal directions would not rotate and 
the two remained coaxial.
That is, if the fabric is characterized
by tensor $\mathbf{a}$ 
(perhaps chosen from the list in Table~\ref{tab:Measures}),
then $(\boldsymbol{\mathbf{a}:\sigma})^{2}=(\mathbf{a:a})(\boldsymbol{\sigma:\sigma})$, 
$(\boldsymbol{\mathbf{\dot{a}}:\sigma})^{2}=(\mathbf{\dot{a}:\dot{a}})(\boldsymbol{\sigma:\sigma})$,
and 
$(\mathbf{a}:\dot{\boldsymbol{\sigma}})^{2}=(\mathbf{a:a})(\boldsymbol{\dot{\sigma}:\dot{\sigma}})$,
as in \cite{Pietruszczak:2001a}. 
The absence of shearing strains
also obviates the need to consider corotational stress rates or increments.
\par
In the study, we exploited a particular advantage of DEM simulations
for exploring material behavior:
once a DEM assembly had been created
and loaded to some initial strain, the precise configuration $\mathcal{S}$
at that instant (particle positions, contact forces, contact force
history, etc.) could be stored and reused with subsequent loading sequences
of almost unlimited variety, all beginning from the same stored configuration
(for example, \cite{Bardet:1994c,Calvette:2003a}).
Following Eq.~(\ref{eq:Darve}), six incremental
tests (three loading and three unloading increments) were begun from
the assembly
configurations at several strains $\varepsilon_{11}$ during the initial
monotonic biaxial loading.
Each set of six incremental oedometric 
(uniaxial compression/extension) tests allow
measurement of eighteen material properties for the three cases $i=1,2,3$
with $i\neq j\neq k$:
\begin{align}
  \DIF\varepsilon_{ii}>0,\; \DIF\varepsilon_{jj}=\DIF\varepsilon_{kk}=0\Rightarrow
  &\begin{cases}
  O_{i}^{+}=\displaystyle{\partial\sigma_{ii}}/{\partial\varepsilon_{ii}}\\
  K_{i}^{j,+}=\displaystyle{\partial\sigma_{jj}}/{\partial\sigma_{ii}}\\
  K_{i}^{k,+}=\displaystyle{\partial\sigma_{kk}}/{\partial\sigma_{ii}}
  \end{cases}\label{eq:Oplus}\\
  \DIF\varepsilon_{ii}<0,\; \DIF\varepsilon_{jj}=\DIF\varepsilon_{kk}=0\Rightarrow
  &\begin{cases}
  O_{i}^{-}=\displaystyle{\partial\sigma_{ii}}/{\partial\varepsilon_{ii}}\\
  K_{i}^{j,-}=\displaystyle{\partial\sigma_{jj}}/{\partial\sigma_{ii}}\\
  K_{i}^{k,-}=\displaystyle{\partial\sigma_{kk}}/{\partial\sigma_{ii}}
  \end{cases}\label{eq:Ominus}
\end{align}
where the $O$ and $K$ are generalized oedometric stiffness moduli and
lateral pressure coefficients. 
Darve and Roguiez \cite{Darve:1999b}
present an octo-linear hypoplastic framework for orthotropic loadings
in which the incremental stress is given as
\begin{equation}
  \begin{bmatrix}\DIF\sigma_{11}\\
  \DIF\sigma_{22}\\
  \DIF\sigma_{33}
  \end{bmatrix}=\mathbf{C}\begin{bmatrix}\DIF\varepsilon_{11}\\
  \DIF\varepsilon_{22}\\
  \DIF\varepsilon_{33}
  \end{bmatrix}+\mathbf{D}\begin{bmatrix}|\DIF\varepsilon_{11}|\\
  |\DIF\varepsilon_{22}|\\
  |\DIF\varepsilon_{33}|
  \end{bmatrix}\label{eq:decomp}
\end{equation}
where matrices $\mathbf{C}$ and $\mathbf{D}$ are the incrementally linear
and incrementally non-linear stiffnesses, defined as
\begin{equation}
\mathbf{C}=\frac{1}{2}\left(\mathbf{Q}^{+}+\mathbf{Q}^{-}\right)
\quad\text{and}\quad%
\mathbf{D}=\frac{1}{2}\left(\mathbf{Q}^{+}-\mathbf{Q}^{-}\right)
\end{equation}
with
\begin{equation}
  \mathbf{Q}^{+}=
  \begin{bmatrix}
  O_{1}^{+} & K_{2}^{1+}O_{2}^{+} & K_{3}^{1+}O_{3}^{+}\\
  K_{1}^{2+}O_{1}^{+} & O_{2}^{+} & K_{3}^{2+}O_{3}^{+}\\
  K_{1}^{3+}O_{1}^{+} & K_{2}^{3+}O_{2}^{+} & O_{3}^{+}
  \end{bmatrix}
\end{equation}
and with the $\mathbf{Q}^{-}$ matrix defined in a similar way,
but with the negative
``$-$'' moduli and coefficients of Eq.~(\ref{eq:Ominus}).
\par
\begin{figure}
  \centering
  \includegraphics{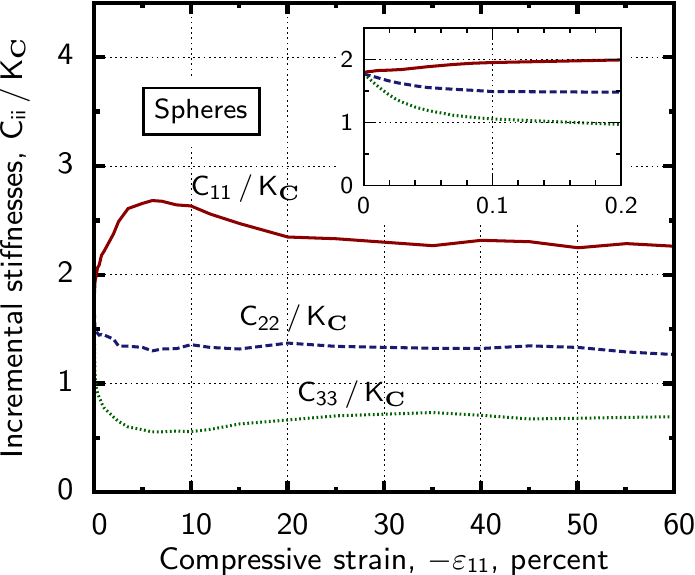}
  \caption{Evolution of incremental
           directional stiffnesses of the sphere assembly.
           Initial compressive loading is in the $x_{1}$ direction.
           Inset plots detail the small-strain stiffness.
  \label{fig:C123}}
\end{figure}

The additive decomposition in
Eq.~(\ref{eq:decomp}) does not expressly
concern elastic and plastic increments:
the linear response $\mathbf{C}$ simply
gives the \emph{\emph{average}} of the loading and unloading stiffnesses;
whereas,
$\mathbf{D}$ is its non-linear hypoplastic complement.
We consider stiffness
$\mathbf{C}$ as more clearly reflective of the anisotropy of the
average bulk stiffness response,
as it can identify differences in the average stiffnesses for
directions $x_{1}$, $x_{2}$, and $x_{3}$.
Figure~\ref{fig:C123} shows the stiffness evolution for the assembly of
spheres when loaded in biaxial compression with constant mean stress.
The directional moduli $C_{11}$, $C_{22}$, and $C_{33}$ have been divided by
the linear bulk modulus $K_{\mathbf{C}}$ (i.e. the average of the
loading and unloading bulk moduli), which is simply equal to the average of
the nine terms of matrix $\mathbf{C}$.
Although the slope of a conventional
stress-strain plot (as in Fig.~\ref{fig:Stress_strain}a)
is greatly reduced during loading, becoming nearly zero beyond the peak-stress
state, the linear $C_{ii}$ moduli are seen to change, with
$C_{11}$ increasing and $C_{33}$ decreasing, but the assembly also retained
stiffness integrity throughout the loading process: the
average loading-unloading moduli were altered, but were not
fully degraded, by the loading.
That is, the granular assembly maintained a load-bearing network of
contacts that continued to provide stiffness,
even as the stress reached a peak and eventually attained a steady-state, 
zero-change condition.
The figure does indicate a developing anisotropy, suggesting that the
load-bearing contact network conferred greater stiffness in the direction
of compressive loading (stiffness $C_{11}$),
while reducing stiffness in the direction of
extension (stiffness $C_{33}$).
Stiffness evolution
$C_{22}$ in the intermediate, zero-strain $x_{2}$ direction
follows an intermediate trend.
\par
\begin{figure}
  \centering
  \mbox{%
    \subfloat[Across directions $x_{1}$--$x_{3}$]{\includegraphics{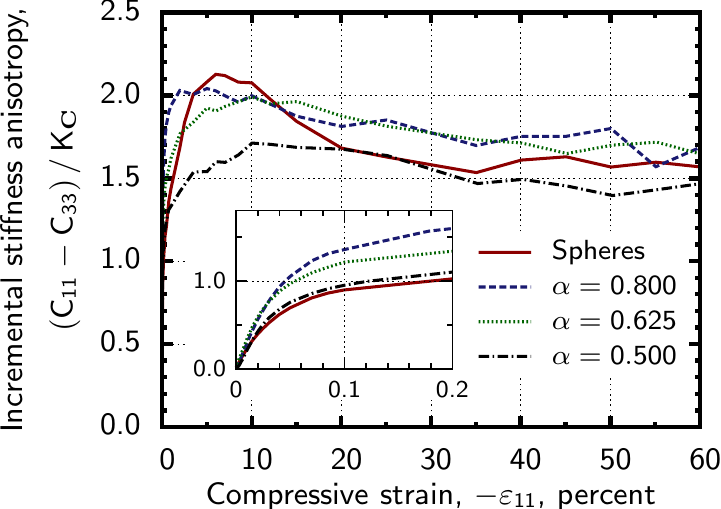}}
    \quad
    \subfloat[Across directions $x_{2}$--$x_{3}$]{\includegraphics{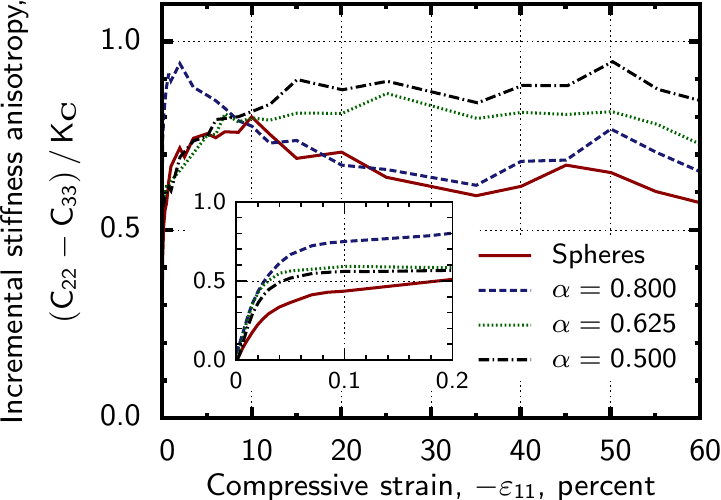}}
  }
  \caption{Anisotropy in the incremental linear stiffness $\mathbf{C}$ 
           of four particle shapes:
           (a) deviatoric anisotropy across the $x_{1}$--$x_{3}$ directions,
           and (b) intermediate deviatoric anisotropy across
           the $x_{2}$--$x_{3}$ directions.
           The stiffness deviator is normalized with respect
           to the average bulk modulus $K_{\mathbf{C}}$.
           Inset plots detail the small-strain stiffness.
           \label{fig:A}}
\end{figure}

The evolution of stiffness anisotropy is more directly measured by
the stiffness differences $C_{11}-C_{33}$ 
and $C_{22}-C_{33}$ (Fig.~\ref{fig:A}).
These measures of stiffness anisotropy are the complements of the stress
anisotropies shown in Figs.~\ref{fig:Stress_strain}a
and~\ref{fig:Stress_strain}b.
Although the numeric
values of the anisotropies of stiffness
and stress
do differ, the trends
are similar across the primary ($x_{1}$--$x_{3}$) 
and intermediate ($x_{2}$--$x_{3}$) directions:
a rapid rise in stiffness (strength), attaining a peak
stiffness (peak strength) at strains of 2--10\%,
followed by a softening (weakening) at larger strains.
At large strains, the deviator stress ratios
across directions $x_{1}$--$x_{3}$
are in the range 0.8--1.3 for the
four particle shapes (Fig.~\ref{fig:Stress_strain}a),
whereas the stiffness difference ratios are 1.4--2.1 (Fig.~\ref{fig:A}a).
At small strains, shown in the insets of
Figs.~\ref{fig:Stress_strain} and~\ref{fig:A},
both deviatoric stress and
stiffness anisotropy increase at the start of loading, although
deviatoric stress
increases with strain more steeply than stiffness anisotropy.
The same trends are apparent across the intermediate directions
$x_{2}$--$x_{3}$ (Figs.~\ref{fig:Stress_strain}b and~\ref{fig:A}b).
\par
These similarities of stress and stiffness anisotropies attest to
stress and stiffness having a common origin in the mechanical interactions
of particles at their contacts.
In Section~\ref{sec:contacts}, we found that anisotropy of
the mixed-fabric strong-contact
tensor,
$\overline{\mathbf{H}}^{\text{c-strong}}$,
correlated most closely with deviatoric stress.
However, we found that
the mixed-fabric contact
tensor among \emph{all} contacts,
$\overline{\mathbf{H}}^{\text{c}}$, most closely
correlates with anisotropy of the incremental stiffness.
The evolution of this fabric measure is shown in
Fig.~\ref{fig:Fcbar_Fcstrongbar}b.
The Pearson coefficient of
the stiffness and fabric anisotropies,
$C_{11}-C_{33}$ and
$\overline{H}^{\text{c}}_{11}-\overline{H}^{\text{c}}_{33}$,
was an average of 0.985 among the four particle shapes,
and the corresponding average correlation for the $x_{2}$--$x_{3}$
anisotropies was 0.981.
Nearly the same correlation was found across all particle shapes,
and $(C_{11}-C_{33})/K_{\mathbf{C}}$ was consistently about 4.1 times greater
than
$(\overline{H}^{\text{c}}_{11}-\overline{H}^{\text{c}}_{33})/\text{tr}(\overline{H}^{\text{c}}_{11}-\overline{H}^{\text{c}}_{33})$
across all strains and all particle shapes.
To summarize, stress is most closely associated with
the fabric of the most heavily loading contacts (the strong-contact network);
whereas, stiffness is most closely correlated with the fabric of 
\emph{all} contacts.
Finally, we note that
an assembly's stiffness is \emph{not} closely
correlated with the orientations of the particle bodies or of the particle
surfaces:
plots of $\overline{\mathbf{J}}^{\text{p}}$,
$\overline{\mathbf{I}}^{\text{s}}$, and $\overline{\mathbf{S}}^{\text{s}}$
(Figs.~\ref{fig:J_bar_p} and \ref{fig:Surfaces_vs_strain}) 
are quite different than those of the
stiffness in Fig.~\ref{fig:A}.
\section{Effective Permeability}\label{sec:permeability}
The particles that form the solid soil
skeleton are often assumed to be impermeable in a time
scale important for most engineering applications.
For this case,
the hydraulic properties of the
granular assemblies are dictated by the geometry of the void space
among the solid grains.
As a result, the effective permeability tensor of a grain assembly
is isotropic if and only if
the micro-structural pore geometry is isotropic.
As was seen in Section~\ref{sec:voids} for cohesionless granular
media, the pore geometry evolves when subjected to external loading.
While continuum-based numerical models, such as
\cite{Sun2013stabilized,Sun2014modeling},
often employ the size of the void space to predict permeability,
the anisotropy of
the effective permeability is often neglected. Certainly, this treatment
may lead to considerable errors in the hydro-mechanical responses if the
eigenvalues of the permeability tensor are significantly different.
\par
\begin{figure}
  \centering
  \mbox{%
    \subfloat[$\alpha=0.500, -\epsilon_{11} = 60\% $]
      {\includegraphics[width=0.465\textwidth]
      {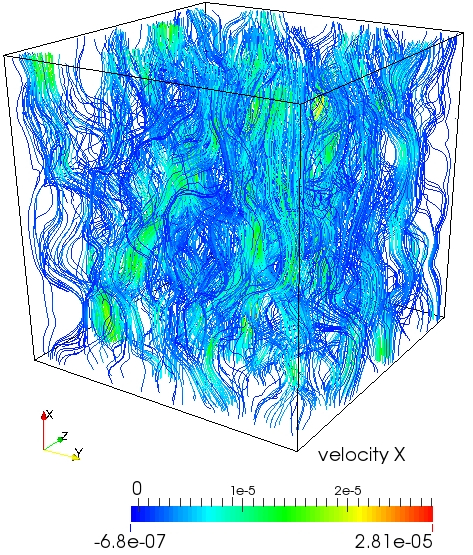}}
    \quad
    \subfloat[$\alpha=0.800, -\epsilon_{11} = 60\% $]
      {\includegraphics[width=0.485\textwidth]
      {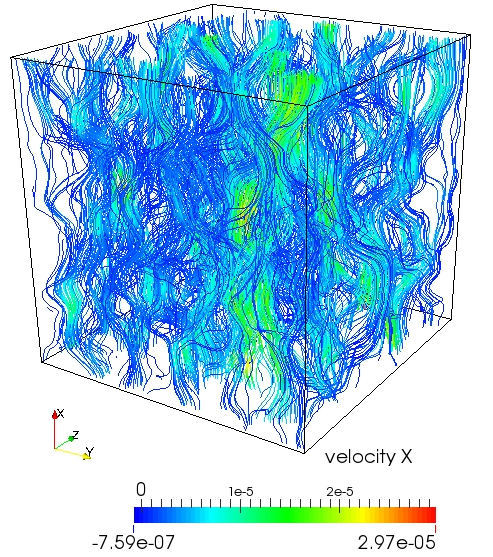}}
  }
  \caption{Streamlines from lattice Boltzmann flow simulations performed on assemblies with $\alpha = 0.500$ and $0.800$
  at $60\%$ shear strain.
           \label{fig:microflow}}
\end{figure}

In this study,
we analyzed the evolution of permeability anisotropy
by recording the positions of all
grains in the assembly at different strains.
As a result, the configuration of the pore space
can be reconstructed and subsequently
converted into binary images
(Fig.~\ref{fig:Digitized}, also \cite{Sun2013}).
To measure effective permeability of a fully saturated porous media,
one can apply a pore pressure
gradient along a basis direction and determine the resultant
fluid filtration velocity
from pore-scale hydrodynamic simulations.
The effective permeability tensor
$\mathbf{K}$ are obtained according to Darcy's law,
\begin{equation}
  \label{eq:permeability}
  k_{ij} = -\frac{\mu^{v}}{p_{,j}} \frac{1}{V_{\Omega}} \int_{\Omega}
  v_{i}(\vec{x}) \DIF\Omega 
\end{equation}
where $\mu^{v}$ is the kinematic viscosity of the fluid occupying the spatial domain of the porous
medium $\Omega$.
The procedure we used to obtain the components of the
effective permeability tensor $k_{ij}$ from Lattice Boltzmann simulation
is as follows.
First, we assumed that the effective permeability
tensor $k_{ij}$ is symmetric and positive definite.
We then determined the diagonal components of the
effective permeability tensor $k_{ii}$
by three hydrodynamics simulations with imposed pressure gradient
on two opposite sides orthogonal to the flow direction and a no-flow boundary condition on the four remaining side faces.
Figure \ref{fig:microflow} shows flow velocity
streamlines obtained from lattice Boltzmann simulations
performed on two deformed assemblies with
grain shapes $\alpha = 0.500$ and $0.800$.
\par
After determining the diagonal components of the effective permeability tensor,
we replaced the no-slip boundary conditions with slip natural
boundary conditions and conducted three additional hydrodynamics simulations,
one for each orthogonal axis.
Since the effective permeability tensor is assumed to be symmetric
and the diagonal components are known, there are three
unknown off-diagonal components that remained to be solved.
To solve the off-diagonal component, we first expanded
Darcy's law,
\begin{equation}
\label{eq:kij1} v_{1} = \frac{1}{\mu^{v}} ( k_{11} \partial p / \partial x_{1} +  k_{12} \partial p / \partial x_{2} + k_{13} \partial p / \partial x_{3})
\end{equation}
\begin{equation}
\label{eq:kij2} v_{2} = \frac{1}{\mu^{v}} ( k_{12} \partial p / \partial x_{1} +  k_{22} \partial p / \partial x_{2} + k_{23} \partial p / \partial x_{3})
\end{equation}
\begin{equation}
\label{eq:kij3} v_{3} = \frac{1}{\mu^{v}} ( k_{13} \partial p / \partial x_{1} +  k_{32} \partial p / \partial x_{2} + k_{33} \partial p / \partial x_{3})
\end{equation}
Putting the known terms on the right sides leads to the system
\begin{equation}
\label{eq:kijsolver}
\left[\begin{array}{ccc} \partial p / \partial x_{2} & \partial p / \partial x_{3} & 0  \\ \partial p / \partial x_{1} & 0 & \partial p / \partial x_{3}  \\ 0 & \partial p / \partial x_{1} & \partial p / \partial x_{2} \end{array}\right] \left[\begin{array}{c} k_{12} \\ k_{13} \\ k_{23} \end{array}\right] =\left[\begin{array}{c} -\mu^{v} v_{1} - k_{11} \partial p / \partial x_{1} \\ -\mu^{v} v_{2} - k_{22} \partial p / \partial x_{2}  \\ -\mu^{v} v_{3} - k_{33} \partial p / \partial x_{3}   \end{array}\right]
\end{equation}
By solving the inverse problem described in Eq.~\eqref{eq:kijsolver}
with the
numerical simulations results from pore-scale simulations,
we obtained the remaining
off-diagonal components of the effective permeability tensor.
In this study, we used the
lattice Boltzmann (LB) method to conduct the pore-scale flow simulations. 
For brevity, we omit description of
the lattice Boltzmann method, and interested readers
are referred to \cite{White2006,Sun2011,Sun2011a,Sun2013} for details.
\par
\begin{figure}
  \centering
  \mbox{%
    \subfloat[Anisotropy across directions $x_{1}$--$x_{3}$]
      {\includegraphics[width=0.485\textwidth]{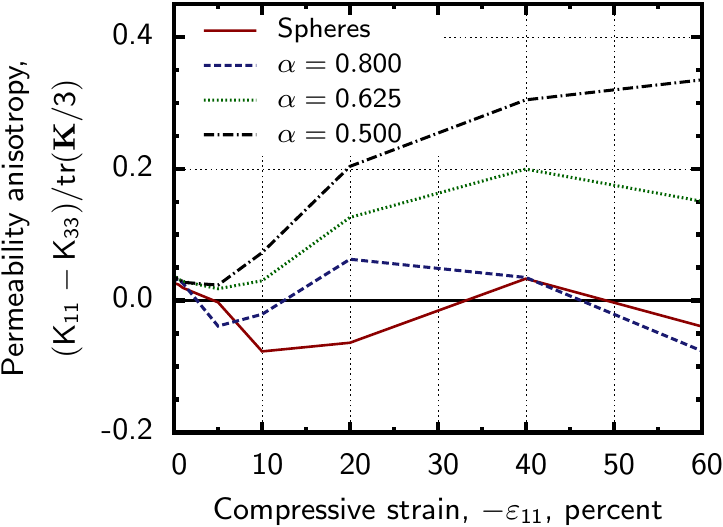}}
    \quad
    \subfloat[Anisotropy across directions $x_{2}$--$x_{3}$]
      {\includegraphics[width=0.485\textwidth]{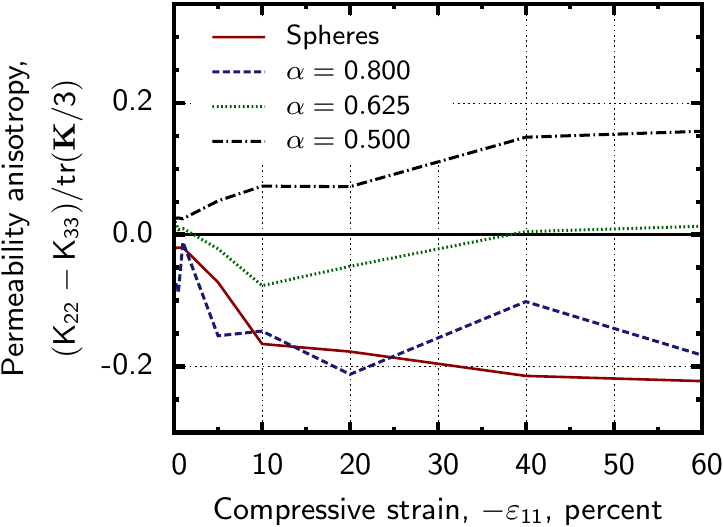}}
  }
  \caption{Induced anisotropy of the effective permeability
           $\mathbf{K}$ for four particle shapes.
           \label{fig:K_vs_strain}}
\end{figure}

Figure \ref{fig:K_vs_strain} shows induced anisotropies in
the effective permeability tensors $\mathbf{K}$ for the four assemblies
during biaxial compression, expressed as
differences among
diagonal components of the effective permeability
tensor, divided by its trace.
%
Differences in the permeabilities of the assemblies of
spheres and of the flatter particles
are apparent.
During early stages of biaxial compression, 
grain assemblies composed of spheres and the most rotund
ovoids have lower permeability in the 
$x_{1}$ direction (the direction of compressive strain) than
in the $x_{3}$ direction (of extension), with
$K_{11} - K_{33} \leq 0$.
Assemblies
composed of the flatter ovoids ($\alpha = 0.625$ and $0.500$),
however,
do not exhibit this trend, and 
compressive strain in the $x_{1}$ direction 
induces a permeability in this direction, $K_{11}$,
that is higher than that in the extensional $x_{3}$ direction,
$K_{33}$
(Fig.~\ref{fig:K_vs_strain}a).
\par
Comparing these trends in the anisotropy of permeability
with anisotropies of the various fabric measures,
we see little correlation between permeability and
the orientations of the particle bodies, of the particles'
surfaces, or of the particles' contacts.
That is,
the plots of $\overline{\mathbf{J}}^{\text{p}}$,
$\overline{\mathbf{I}}^{\text{s}}$, $\overline{\mathbf{S}}^{\text{s}}$,
$\overline{\mathbf{F}}^{\text{c}}$, etc.
(Figs.~\ref{fig:J_bar_p}, \ref{fig:Surfaces_vs_strain},
and~\ref{fig:Fcbar_Fcstrongbar})
are quite different from those of the
permeability in Fig.~\ref{fig:K_vs_strain}.
We do see, however, similarities between
the anisotropies of permeability and those of
the median free path and the median radial breadth of
the void space (see Figs.~\ref{fig:Lambda_vs_strain}
and~\ref{fig:LLambda_vs_strain}).
Anisotropy in the permeability $\mathbf{K}$
is negatively correlated with the median free path
matrix $\overline{\mathbf{L}}^{\text{v}}$ and is
positively correlated with the median radial breadth
matrix $\overline{\mathbf{R}}^{\text{v}}$.
These trends are apparent for anisotropies across both the
$x_{1}$--$x_{3}$ and $x_{2}$--$x_{3}$ directions.
These trends suggest two competing influences on
the effective permeability.
On the other hand,
a larger median free path in a particular direction indicates
a reduced tortuosity in this direction, which should increase
the directional permeability: a trend that is at variance
with the counter-correlated trends in Figs.~\ref{fig:Lambda_vs_strain}
and \ref{fig:K_vs_strain}.
A larger median radial breadth in a particular direction
is consistent with a larger hydraulic radius for flow in
this direction, and anisotropies in the median radial breadth
$\overline{\mathbf{R}}^{\text{v}}$ and effective permeability
$\mathbf{K}$ should be correlated, which is in accord
with the positively-correlated
trends of Figs.~\ref{fig:LLambda_vs_strain}
and \ref{fig:K_vs_strain}.
The numerical experiments indicate that
change in the directional hydraulic radii is the more dominant mechanism
in influencing the induced anisotropy of the effective permeability. 
This result is probably attributed to the fact that the void spaces
of all four assemblies are highly interconnected and of relatively
high porosity.
\section{Conclusion}
Thirteen measures of fabric are 
arranged in four categories, depending upon the
object of interest:
the particle bodies, the particle surfaces, the contacts,
and the voids.
The orientations of the particle bodies and their surfaces are fairly easy to
measure, and their induced anisotropies follow similar trends 
during monotonic biaxial
compression.
Anisotropies of these measures increase with loading, but their change
lags changes in the bulk stress, and they continue to change even after
stress and volume have nearly attained steady values; in particular,
non-spherical particles continue
to be reoriented at strains greater than 60\%.
Although they are easiest to measure,
the average orientations of particle bodies and their
surfaces are poor predictors
of stress, incremental stiffness, and effective permeability.
The mechanical response,
stress and stiffness, are more closely associated with contact orientation.
A mixed tensor, involving both contact and branch vector orientations,
is most closely correlated with the stress and stiffness.
Stress is closely correlated with the most heavily loaded contacts
within an assembly (the strong-contact network);
whereas, the average orientation of all contacts is most closely
correlated with the bulk incremental stiffness.
In short, tensor $\overline{\mathbf{H}}{}^{\,\text{c-strong}}$
is the preferred fabric measure for stress, and
tensor $\overline{\mathbf{H}}{}^{\,\text{c}}$
is the preferred fabric measure for incremental stiffness.
Two principal measures of pore anisotropy were investigated
in regard to the effective permeability:
one related to the directional median free path
(a counter-measure of tortuosity), and the
other related to the directional median radial breadth
(a measure of hydraulic radius).
The preferred measure for effective permeability is the matrix
of the median radial breadths of the void space,
$\overline{\mathbf{R}}^{\text{v}}$, as it correlates
closely with permeability.
\begin{acknowledgements}
This research is partially supported by
the Earth Materials and Processes program at
the US Army Research Office under grant
contract W911NF-14-1-0658 and the Provosts Grants
Program for Junior Faculty who Contribute to the
Diversity Goals of the University at Columbia University.
The Tesla K40 used for the lattice Boltzmann simulations
was donated by the NVIDIA Corporation.
These supports are gratefully acknowledged.
\end{acknowledgements}
\bibliographystyle{plain}



\end{document}